\documentclass [12pt,preprint]{aastex} 
\usepackage{epsfig} %
\begin{document} 
\voffset-0.5cm 
\newcommand{\gsim}{\hbox{\rlap{$^>$}$_\sim$}} 
\newcommand{\lsim}{\hbox{\rlap{$^<$}$_\sim$}}

\title{High Energy Phenomena In The Universe}

\author{Arnon Dar\altaffilmark{1}}

\altaffiltext{1}{arnon@physics.technion.ac.il\\
Physics Department, Technion, Haifa 32000, Israel}

\begin{abstract}

Highlights of the 44th Rencontre De Moriond on High Energy Phenomena In 
The Universe which was held in La Thuile, Italy during February 1-8, 2009.
\end{abstract}

\section{Introduction} 

More than 110 talks and 10 posters were presented at the 44th Rencontre De 
Moriond on high energy phenomena in the universe. They reflect the flood 
of new and important results in the fields of cosmic ray astrophysics, 
high energy gamma ray astronomy, high energy neutrino astronomy and the 
search for astrophysical evidence of physics beyond the standard models of 
particle physics, general relativity and cosmology. Unable to cover in a 
short summary all the talks and the new results, I will limit my summary 
and comments to results which were presented and discussed in this 
Rencontre and which to the best of my judgment are the most important and 
fundamental ones.

\section{Ultra-high Energy Cosmic Rays} 

If the ultra-high energy cosmic rays (UHECRs) observed reaching Earth
are extragalactic in origin, 
as suggested by the isotropy of their arrival directions and the lack of 
correlation with the Galactic plane, than inelastic collisions with the 
cosmic background radiation (CBR) and cosmic expansion are expected to 
degrade their energies during their travel from their extragalactic 
sources to Earth. If the UHECRs are protons, pion production 
in collisions with the cosmic microwave background radiation (MBR) 
strongly degrades their energy above an effective threshold of 
$\!\sim\! 5\times 10^{19}$ eV, the so called Greisen-Zatsepin-Kuzmin (GZK) 
threshold~\cite{Gr},\cite{Za} while $e^+e^-$ pair production in collisions 
with the CBR degrades their energy above an effective threshold of 
$\!\sim\!10^{18}$ eV just below the CR ankle at $\!\sim\! 3\times 10^{18}$ 
eV. If the UHECRs are nuclei, nuclear photodissociation in collisions 
with 
the CBR begins to be effective at a slightly lower energy for light 
nuclei and around the GZK threshold energy for iron-like nuclei~\cite{Al}. 
Thus, the suppression of the flux of CR protons above the GZK threshold is 
expected to be accompanied by even a stronger suppression of the flux of 
heavier nuclei.

Early measurements by the Akeno Giant Air Shower Array (AGASA), which 
detects air showers at ground level with scintillators, reported the 
detection of UHECRs above the GZK threshold not showing the expected GZK 
suppression~\cite{Ta} but showing strong clustering in their arrival 
direction. These led to variety of interpretations including speculations 
on physics beyond the standard particle physics model and on violation of 
Lorentz invariance and special relativity. However, later results from the 
High Resolution Fly's Eye (HiRes) experiment~\cite{Abb},\cite{Be}, which 
detect the fluorescence emitted in the air by nitrogen molecules excited 
by the passage of the shower, observed the GZK suppression above the 
expected threshold and did not find a significant unisotropy in the 
arrival directions of UHECRs. The AGASA and HiRes results were based on a 
small number of events and used different techniques. Results from 
measurements of UHECRs by the Pierre Auger Observatory which was 
conceived as a hybrid detector combining the two detection methods and 
covering an area 30 times bigger than that of AGASA, that were obtained 
during its construction confirmed the GZK suppression above the expected 
threshold~\cite{Ab},\cite{Sc} and appeared to indicate that UHECRs above 
the GZK threshold arrive from nearby active galactic 
nuclei~(9),\cite{PAO2}.

\begin{figure}[]
\centering
\vbox{
\hskip -2.5cm 
\epsfig{file=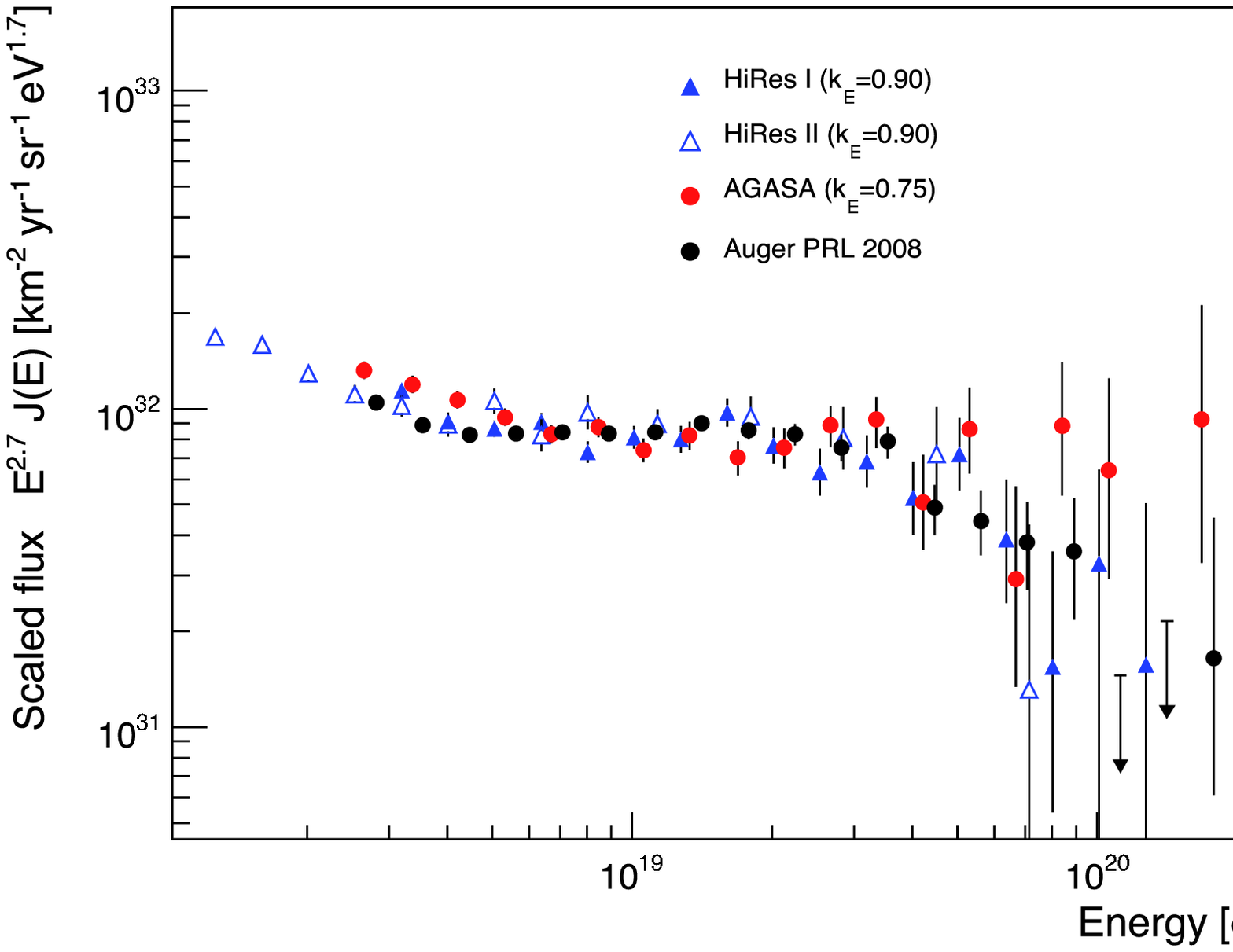,width=14.0cm,height=9.0cm}
\epsfig{file=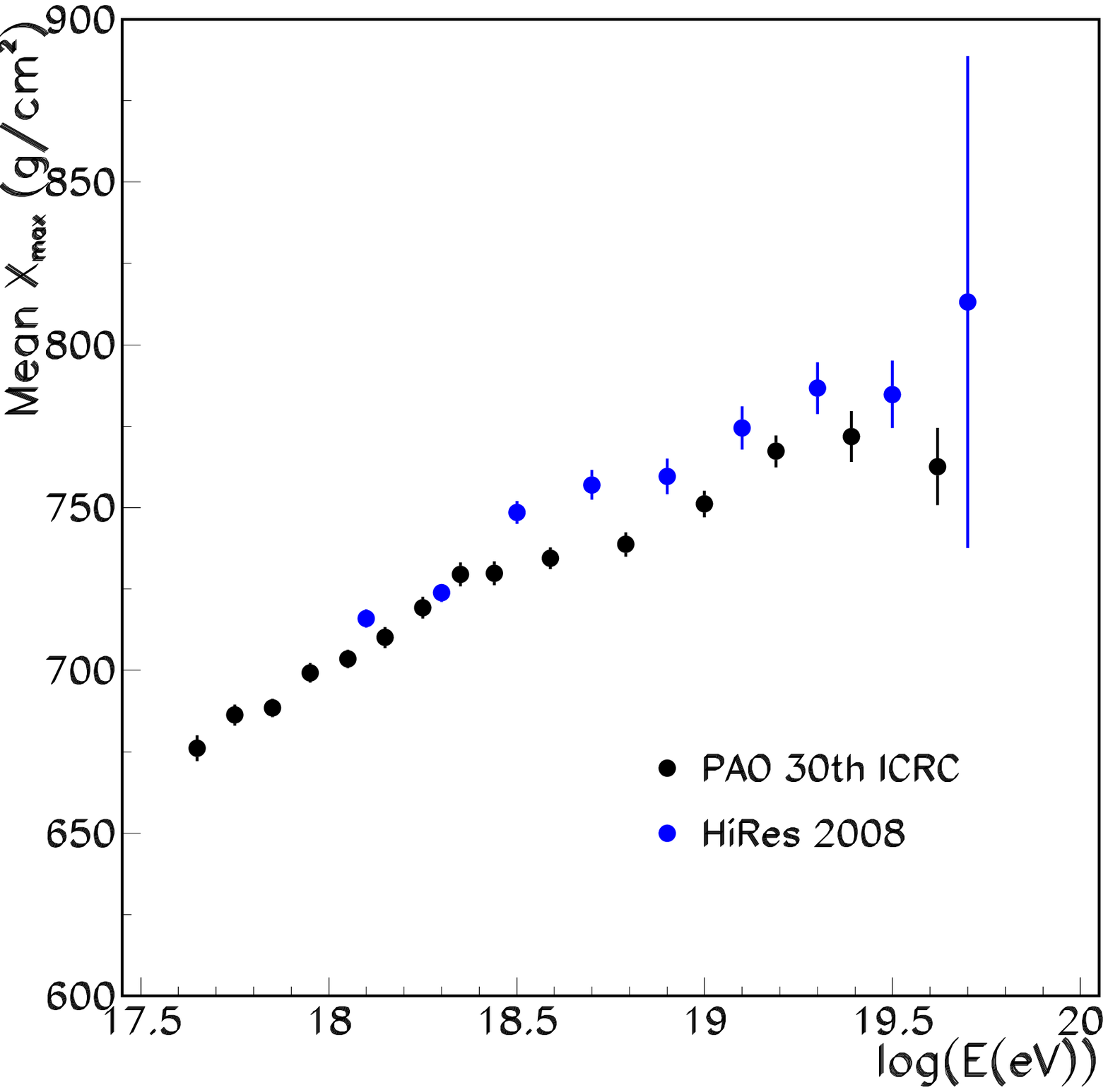,width=12.0cm,height=9.0cm}
}
\caption{{\bf Top (a):} Comparison between the spectra of UHECRs 
multiplied by $E^{2.69}$ measured by PAO, HiRes (with energy rescaled by a 
factor 0.9) and AGASA (with energy rescaled by a factor 0.7). The PAO and 
HiRes data are consistent and show the expected GZK suppression above 
$4\times 10^{19}$ eV. {\bf (Bottom (b):} Comparison between the mean depth 
of shower maximum of UHECRs as measured by HiRes and by PAO.} 
\label{UHECRs12} 
\end{figure}

\begin{figure}[] 
\centering
\vspace{-8cm}
\vbox{ 
\epsfig{file=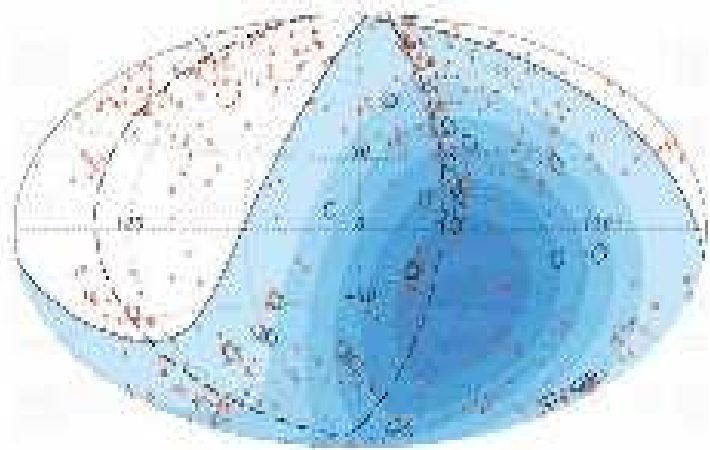,width=14.cm,height=18.0cm}  
\vspace{-2.cm}
\epsfig{file=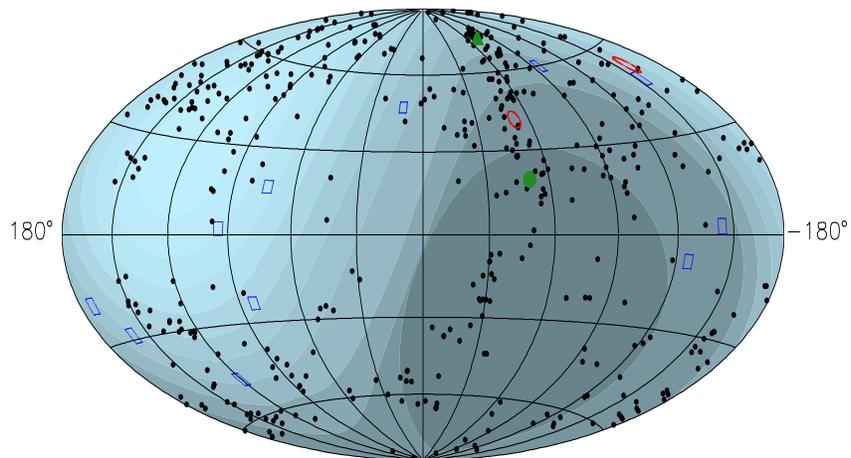,width=12.0cm,height=9.0cm} 
}
\caption{{\bf Top (a):} The arrival directions of UHECRs with energy above 
57 
EeV, measured by PAO and plotted as circles with an angular radius of 
$3.2^o$ centered on their arrival direction on a sky map (Galactic 
coordinates) of AGNs within 71 Mpc from Earth. Colors indicate equal 
exposure.
{\bf Bottom (b):} 
The arrival directions of UHECRs with energy above 57 EeV, measured by PAO 
and plotted as circles with an angular radius of $3.2^o$ centered on their 
arrival direction  on a sky map (Galactic coordinates) of AGNs within 
71 Mpc from Earth. Colors indicate equal exposure.} 
\label{UHECRs34}
\end{figure}

The fast falling spectrum of the ultra-high energy cosmic rays (UHECRs), 
up to energies of about $10^{20}$ eV where the CR flux is of the order of 
1 particle per km$^2$ per a couple of centuries, their arrival directions 
and their composition have now been measured by HiRes~\cite{Be} and by the 
PAO~\cite{Sc},\cite{Ma},\cite{Kn} with sizable statistics (roughly 
twice and four times, respectively, the exposure of AGASA). The main 
results can be summarized as follows:

\begin{itemize}

\item{} {\bf GZK Suppression Confirmed:} 
Allowing for 10\% adjustment 
in the CR energies inferred either by HiRes or PAO from the 
flourescence light emitted by air molecules excited by the CR 
induced atmospheric showers, 
because of a 10\% difference in the adopted flourescence yield in the
showers, the energy spectra of UHECRs measured by both experiments are 
identical (Fig.~\ref{UHECRs12}a) and show the expected GZK 
suppression beyond $\!\sim\! 4\times 10^{19}$ eV, consistent with the 
highest energy CRs being extragalactic protons. (The power law $E^{-2.69}$ 
which fits the PAO spectrum below 40 EeV predicted 163$\pm$3 events above 
40 EeV and 35$\pm$1 above 100 EeV, while 69 events and 1 event were 
observed by PAO, clearly confirming the GZK suppression).

\item{}{\bf Composition:} 
The atmospheric depth (in g/cm$^2$) of shower maximum, $X_{max}$, has been 
used both by HiRes~\cite{Be} and PAO~\cite{Ma},\cite{Sch} to infer the 
composition of UHECRs. Both experiments report  a mixed composition that 
is becoming lighter with energy up to 3 EeV. However, HiRes results 
indicate a light composition all the way up to the GZK threshold around 40 
EeV where it runs out of statistics, whereas PAO results indicate that the 
composition becomes heavier above 3 EeV and more so beyond the GZK 
threshold (Fig.~\ref{UHECRs12}b). These conclusions are valid 
provided that hadron physics does not change above 3 EeV.

\item{}{\bf Isotropy:} 
Below the GZK threshold both the HiRes and the PAO 
CR events are completely consistent with statistical fluctuations of an 
isotropic distribution of arrival directions.

\item{} {\bf UHECRs-AGN correlation:} At energies above the GZK threshold 
only CRs from nearby sources can reach Earth. If they are not deflected 
much by the intergalactic and Galactic magnetic fields, their arrival 
directions should point back to their sources, opening the window to UHECR 
astronomy. The evolution with energy of the distribution of arrival 
directions of UHECRs measured by PAO shows a sharp transition from 
isotropy to anisotropy beyond the GZK threshold. The arrival directions of 
UHECRs with energy above 57 EeV show a correlation on angular scales of 
less than $6^o$ with the sky positions of AGNs within 71 Mpc, which are 
concentrated near the supergalactic plane. Intrinsic (catalog independent) 
properties of these events, such as their auto-correlation function, show 
a clear departure from isotropy in a large angular range~\cite{Kn}. The 
correlation/unisotropy observed by PAO was not confirmed by HiRes which 
reported~\cite{Be} lack of arrival-direction correlation of their 
highest energy events with local AGNs (in the Northern Hemisphere). PAO 
found that out of their 27 UHECRs events with energy above 56 EeV, 20 were 
found to lie within $3.2^o$ of the line of sight to an AGN nearer than 71 
Mpc (Fig.~\ref{UHECRs34}a)
while only 6 were expected to be found by chance from an isotropic 
distribution of arrival directions (the threshold energy, maximal angular 
deviation, and maximal AGN distance were chosen to maximize the 
UHECRs-correlation). HiReS found that 
using the PAO criteria only 2 of their 13 events above 56 EeV correlated 
with AGN (Fig.~\ref{UHECRs34}b), while 3.2 were expected randomly, 
ruling out the correlation at a probability of 83\%. The PAO collaboration 
has stressed that even though the correlation with nearby AGN seems to be 
quite robust in their sample, the angular scale of $\! \sim\!6^o$ does not 
make possible to unambiguously identify the sources and sources which are 
distributed similar to AGNs cannot be excluded as the true sources.

\item{}{\bf UHEGRs:} 
Showers initiated by ultra-high energy gamma rays (UHEGRs) develop 
differently from showers induced by nuclear primaries. Particularly, the 
depth of shower maximum is much larger and the shower is much poorer in 
muons relative to those of CR nuclei. Upper limits on the presence of 
photons in the primary cosmic-ray flux were obtained by PAO; in particular 
a limit of 2\% (at 95\% c.l.) above 10 EeV on the flux of UHEGRs relative 
to UHECRs was derived by PAO~\cite{Sch}. This limit improves previous 
constraints on Lorentz violation parameters by several orders of magnitude 
due to the extreme energy in case of UHEGRs.

\end{itemize}

Although AGN are a natural source of extragalactic UHECRs, the 
directional correlation found by Auger is surprising in many respects. A 
$3.2^o$ deviation is of the order of magnitude of that inflicted on UHECRs 
by the magnetic field of the Galaxy, it would be surprising if 
extragalactic CRs did not encounter intergalactic magnetic fields with 
similar or larger effects. The Veron catalog of AGN is not complete and 
not directionally uniform in its coverage and sensitivity, unlike the 
Auger coverage within its field of view. The Auger correlation is purely 
directional, not investigated case-by-case for the possible effects of AGN 
distance, luminosity, jet direction and radio loudness. The effect of 
distance is obvious, the correlation with luminosity is very plausible. 
Concerning jet-direction, one has to understand how the UHECRs from AGNs 
could be fairly isotropically emitted, given that AGNs produce extremely 
collimated jets, and that they are seen in gamma-rays as very luminous 
blazars only when the jets are pointing in our direction. The proton- and 
electron-acceleration efficiencies of CR sources are presumably 
correlated. The radio loudness is a measure of the number of high energy 
synchrotron-radiating electrons.
The jets of an AGN may accelerate CRs to well above the GZK limit and 
collimate them forward in a cone of aperture $1/\Gamma$ where $\Gamma$ is 
their bulk motion Lorentz factor. But the PAO results suggest a more 
isotropic source, the end lobe of an AGN jet being the obvious 
choice~\cite{Ra}. 
These lobes have radii $R_l$ of a few kpc. They are steadily energized by 
the incoming jet. Traveling in a medium swept up by previous jet 
components, a jet may deposit in its lobe an energy in excess of $10^{60}$ 
erg, emitted by the central black hole during the AGNs active life. An 
equipartition magnetic field $B$ in these end lobes can exceed a milli 
Gauss. The Larmor limit energy for the acceleration of a proton in a lobe 
is then $E_{max}\!\approx\! e\,B\,R_L\!\approx\! 3\times 10^{21}$ eV, well 
above the GZK threshold.
 
However, the PAO UHECRs-AGN correlation is puzzling in other respects. 
E.g., why there are no events from the direction of the Virgo cluster, 
that contains powerful AGN such as M87 at 14 Mpc? Why the maximal 
correlation for UHECRs with E$\!\geq\! 57$ EeV is with AGN at distance 
less than 71 Mpc - Such UHECRs should come from distances up to 200 Mpc 
and not only from less than 71 Mpc.
 
All together, the results from PAO are very important in many respects and 
are pointing towards a potential breakthrough in UHECR and UHEGR 
astronomies, but much more statistics are needed in order to establish 
that. With a main goal of full sky coverage, the Auger Observatory is to 
be completed by a northern site. Current plans aim at a significantly 
$\!\sim\! 7$ times larger array to proceed with UHECR and UHEGR 
astronomies.
  
To reach even larger exposures, dedicated observatories in space which can 
observe UHECR induced atmospheric showers by looking down towards the 
Earth are planned. The Extreme Universe Space Observatory (EUSO) on the 
Japanese Experiment Module (JEM), which will detect fluorescence from 
UHECR 
events within $60^o$ field of view, is being planned for deployment on the 
International Space Station. JEM-EUSO may detect $\!\sim\!1,000$ particles 
above 70 EeV in a three year mission. The Orbiting Wide-Angle Light 
Collectors (OWL) will stereoscopically image fluorescence from UHECRs. 
Such missions may observe a significant fraction of the $\!\sim\!$ 10 
million showers generated in the Earth atmosphere per year by UHECRs with 
energy above the GZK threshold.

\section{Dark Matter}

\subsection{\bf Evidence from cosmic colliders} 

Dark matter is an hypothetical matter that does not emit electromagnetic 
radiation, whose presence has been inferred consistently from 
gravitational effects on visible matter, on light trajectories, on the 
space-time geometry of the universe, on structure formation in the 
universe and on cosmic evolution.
 
The observed phenomena which imply that the universe contains much more 
dark matter than visible matter, include the rotational speeds of 
galaxies, orbital velocities of galaxies in clusters, gravitational 
lensing of background objects by galaxies and galaxy clusters and the 
temperature distribution of hot gas in galaxies and clusters of galaxies. 
Dark matter also plays a central role in structure formation and galaxy 
evolution, and has measurable effects on the anisotropy of the cosmic 
microwave background radiation. At present, the density of ordinary 
baryons and radiation in the universe is estimated about 4\% of the total 
energy density in the universe. About 22\% is thought to be composed of 
dark matter. The remaining 74\% is thought to consist of dark energy, 
distributed diffusely in space.

\begin{figure}[]
\centering
\vspace{-1.cm}
\vbox{
\epsfig{file=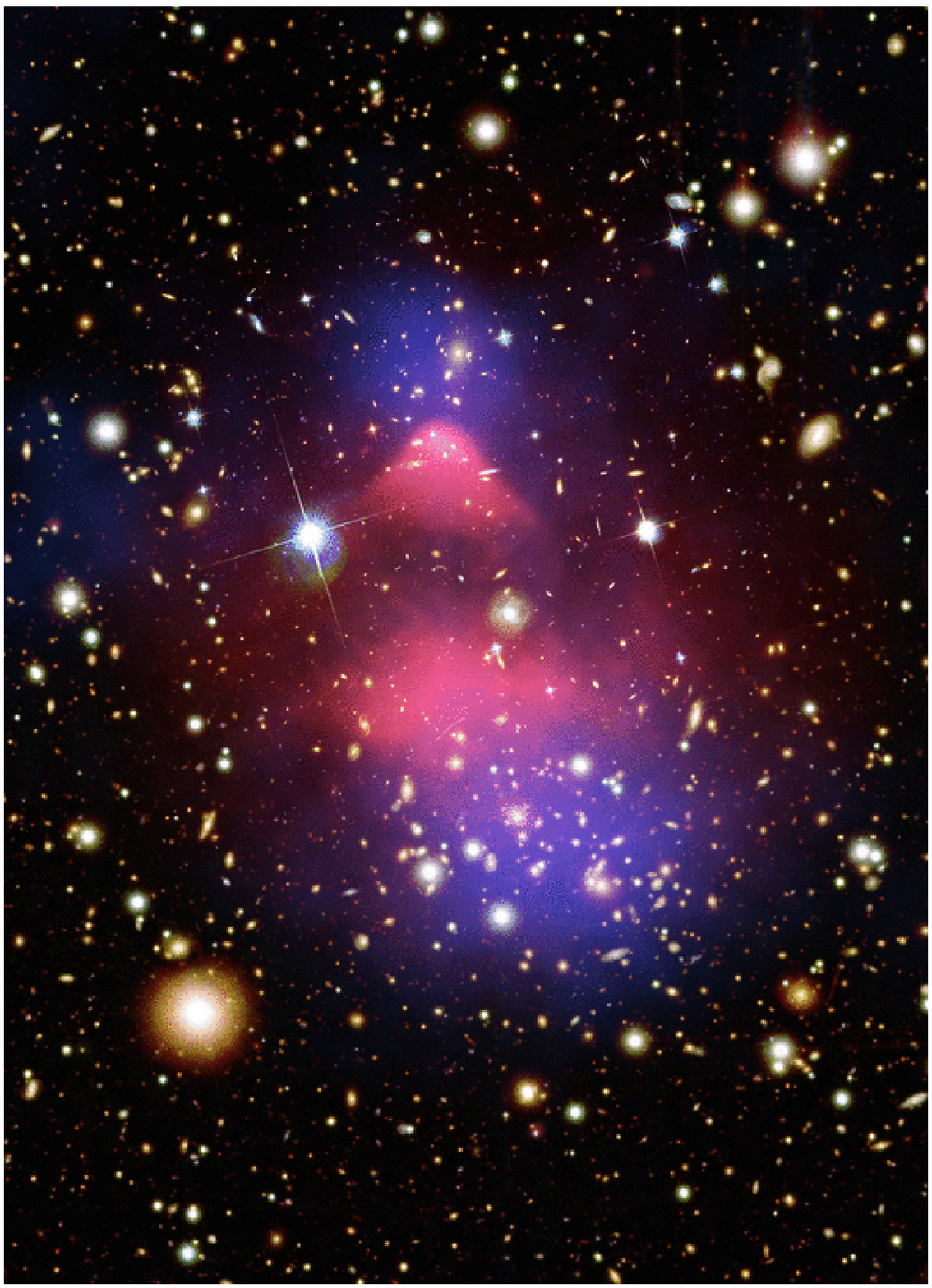,width=14.cm,height=9.cm}
\epsfig{file=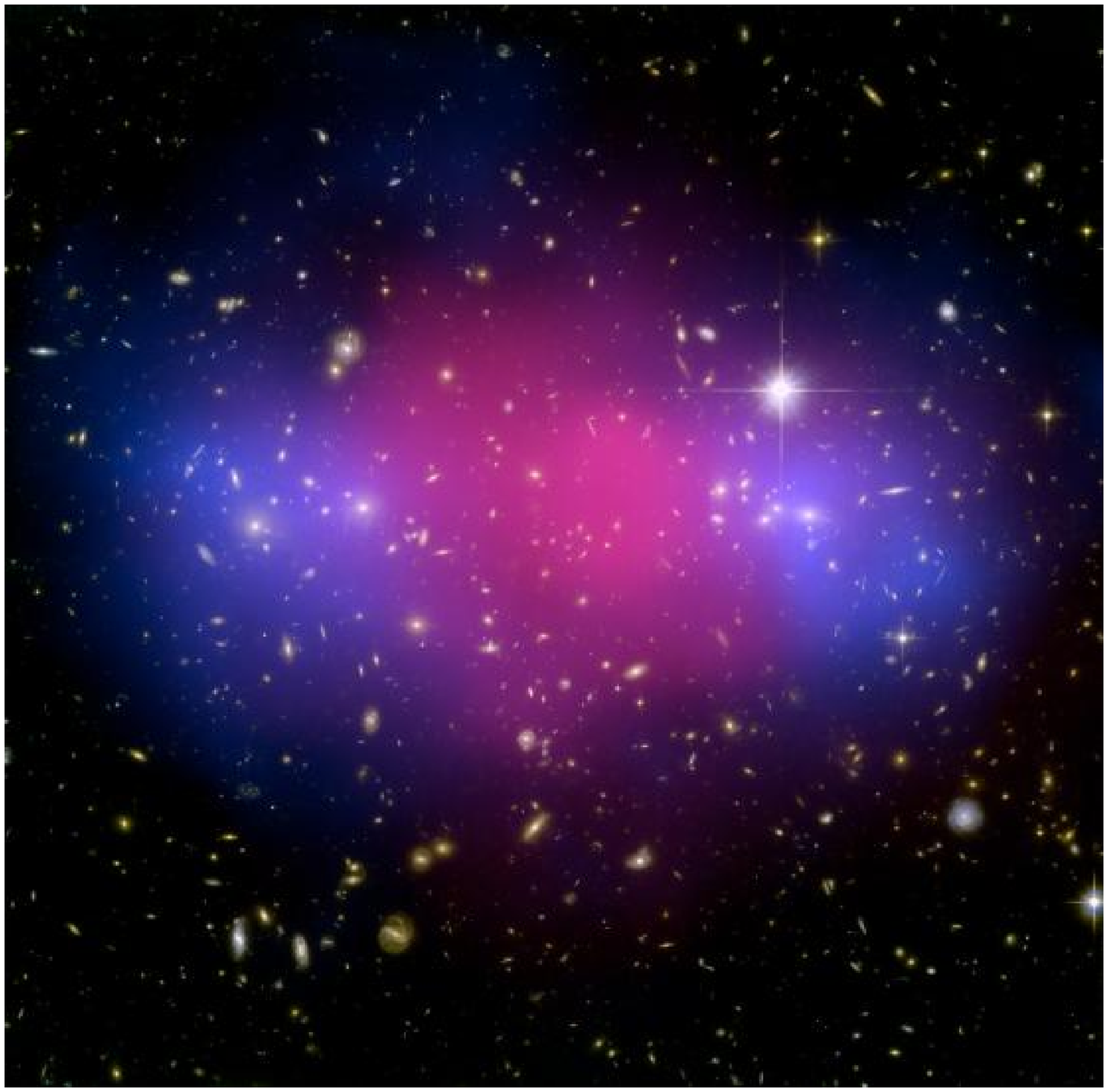,width=12.0cm,height=9.0cm}
}
\caption{Composite images of the bullet cluster 1E 0657-56
({\bf Top (a)}) and the cluster MACS J0025 ({\bf Bottom (b)}). 
Both clusters were formed by a collision of two  galaxy clusters.  
The major components of the clusters  are shown in different colors., 
The galaxies 
whose stars makes them visible in optical light 
are shown in orange and white,
the ionized gas in the clusters which is visible in 
X-rays is shown in pink  and the 
putative dark matter, which dominates their gravitational potential 
and is inferred  through gravitational lensing of background 
galaxies, is shown in blue. After the collision, most of the matter in the 
clusters (in blue) is well separated from most of the normal matter 
(the gas in pink) and moves ahead of it. This separation provides
direct evidence that most of the matter in the 
clusters is dark matter which cannot be represented by modified 
gravity of the cluster gas which contains most of the baryons in  
clusters. Credits
1E0657-56: X-ray NASA/CXC/CfA Optical: NASA/STScI; Magellan/U.Arizona; 
Clowe et al. (2006); Bradac et al. (2006)
MACS J0025.4-1222: X-ray(NASA/CXC/Stanford/S.Allen); Optical/
Lensing(NASA/STScI/UCSB/M.Bradac) Bradac et al. (2008)
}
\label{DM}
\end{figure}

The dark matter hypothesis has generally been the preferred solution to 
the missing mass problems in astronomy and cosmology over alternative 
theories of gravity based on modifications to general relativity which 
have been used to model dark matter observations without invoking dark 
matter~\cite{Beck},\cite{Br}. However, until recently there was no 
conclusive evidence that dark matter really exists. This has changed 
dramatically by X-ray and optical observations of collisions 
between galaxy 
clusters~\cite{Bradac},\cite{Bradacetal1},\cite{Bradacetal2}, 
such as in 1E0657-558 at 
$z$=0.296 (the `Bullet Cluster'), MACS J0025 at $z$=0.586 and A520 at 
$z$=0.201 (the `Cosmic Train Wreck'). In such collisions the clusters' 
galaxies and dark matter halos are affected only by gravity while the 
electromagnetic interactions between the clusters' X-ray emitting ionized 
gas produce an additional drag on the gas. 
Consequently, after the collision the galaxies and their associated dark 
matter halos lead the slower moving  
X-ray emitting gas clouds stripped off from the galaxy clusters, as 
seen in Figs.~\ref{DM}a,\ref{DM}b. 
The galaxies in these Figures were observed from 
the ground with Magellan and from space with the HST, the stripped off 
X-ray emitting gas was mapped with Chandra and the dark matter halos of 
the clusters were mapped by measuring the distortion of the images of 
background galaxies by the deflection of light as it passes the clusters 
dark matter halos. Such observations require that regardless of the form 
of the gravitational force law at large distances and low accelerations, 
the majority of the mass of the system be some form of dark matter. Many 
more cases of cluster collisions will be studied through gravitational 
lensing of background galaxies with a dedicated large telescope such as 
the 8.4m Large Synoptic Survey Telescope (LSSS) which is under design and 
development and scheduled to be commissioned at Cerro Pach¢n (Chile) by 
2017~\cite{Bradac}.

\subsection{\bf Direct and indirect detections ?}

Determining the nature of the dark matter particles is one of the most 
important problems in modern cosmology and particle physics. Both direct 
detection in which the interaction of dark matter particles are observed 
in a detector and indirect detection that looks for the products of dark 
matter annihilation or decay products have been conducted extensively and 
are ongoing. Dark matter detection experiments have ruled out some WIMP 
(Weakly Interacting Massive Particle) and axion models. There are also 
several claims of direct detection of dark matter particles in lab 
experiments such as DAMA/NaI (Dark Matter/Sodium Iodine) in the Gran Sasso 
underground laboratory, and possible detections of astrophysical gamma 
rays, positrons and electrons from dark matter annihilation, by EGRET 
aboard the CGRO, by ATIC and by PAMELA, respectively, but all these are so 
far unconfirmed and difficult to reconcile with the negative results of 
other experiments. In particular:

\subsubsection{\bf The EGRET GeV excess:}

\begin{figure}[]
\centering
\vspace{1.cm}
\vbox{
\epsfig{file=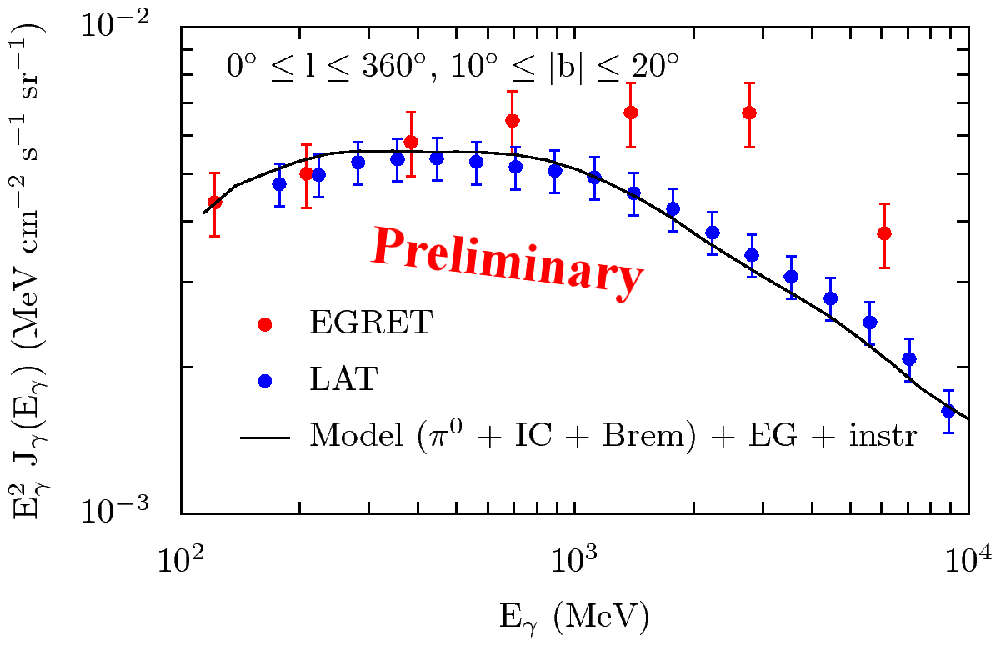,width=12.cm,height=9.0cm}
\epsfig{file=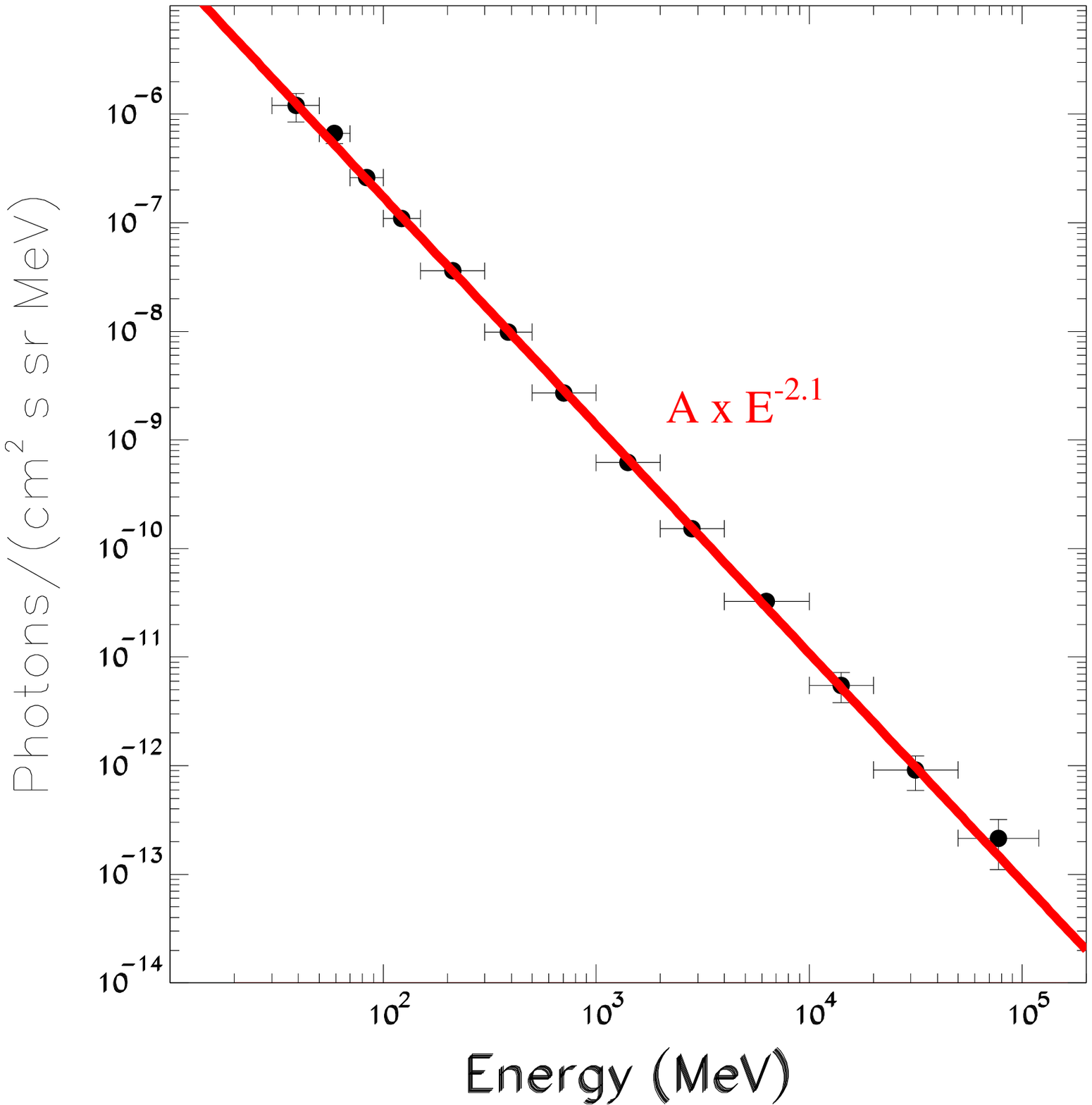,width=13.cm,height=8.cm}
}
\vspace{-0.5cm}
\caption{ {\bf Top (a):} Comparison between the spectra of the diffuse 
gamma ray background radiation at intermediate latitude
which were measured by EGRET~\cite{GeVEGRET} and by  LAT~\cite{GeVLAT}. 
The LAT data do not 
confirm the existence of the EGRET GeV excess and can be fitted by the 
standard model of Galactic cosmic ray electrons and nuclei with densities
normalized to their respective locally observed densities.
{\bf Bottom (b):} The spectrum of the extragalactic gamma ray
background radiation (GBR) which was measured by EGRET~\cite{Sreek} and
is well represented by a single power law,
$dn/dE\!\propto\!E^{-2.10\!\pm\!0.03}$. No  dark matter annihilation/decay 
fingerprints are evident in the EGRET extragalactic GBR. 
}
\label{GBR}
\end{figure}

The spectrum of the diffuse $\gamma$ background radiation (GBR) that was 
measured by EGRET aboard the Compton Gamma Ray Observatory 
showed an excess above 1 GeV in comparison with the 
flux expected from interactions of cosmic ray (CR) nuclei and electrons in the 
Galactic interstellar medium (ISM)~\cite{GeVEGRET}. The origin of
this GeV 
excess has been unknown. Among its suggested origins was annihilation or 
decay of WIMPs~\cite{DMARev}. However, recent measurements with the Large 
Area Telescope (LAT) aboard the Fermi observatory have yielded preliminary 
results~\cite{GeVLAT} which do not show a GeV excess at small Galactic 
latitudes and agree with the flux expected from CR interactions in the 
Galactic ISM (Fig.~\ref{GBR}a). Also, the extragalactic GBR measured by 
EGRET does not show a correponding `GeV excess' (Fig.~\ref{GBR}b).
which would be expected from such dark matter annihilation/decay in 
external galaxies and in the 
IGM. Moreover, by comparing the spectra of gamma-rays around one GeV 
from nearby Galactic pulsars, which were measured by EGRET and LAT, the 
Fermi collaboration confirmed~\cite{GeVLAT} previous 
conclusions~\cite{Stecker} that the origin of the EGRET GeV excess is 
probably instrumental and not a dark matter annihilation/decay signal.

\subsubsection{\bf The ATIC excess:}

The Advanced Thin Ionization Calorimeter (ATIC) experiment aboard balloon 
flights over Antarctica~\cite{ATIC} reported an excess in the flux of CR 
electrons at energies between 300-800 GeV. Several papers suggested that 
this excess in cosmic ray electrons (and positrons) arises from 
annihilation of dark matter particles such as Kaluza-Klein particles with 
a mass of about 620 ${\rm GeV/c^2}$~\cite{DMKZ}). However, in this meeting 
caution was advocated when interpreting cosmic ray electron and positron 
data above a few GeV because of possible proton contamination of the 
measurements and it was pointed out that the ATIC reported data should be 
suspected as the authors did not properly take into account the 
uncertainties associated with a potential hadronic background due to 
particle interactions inside the graphite target on top of the 
detector~\cite{Schub}.

\begin{figure}[]
\centering
\vbox{
\epsfig{file=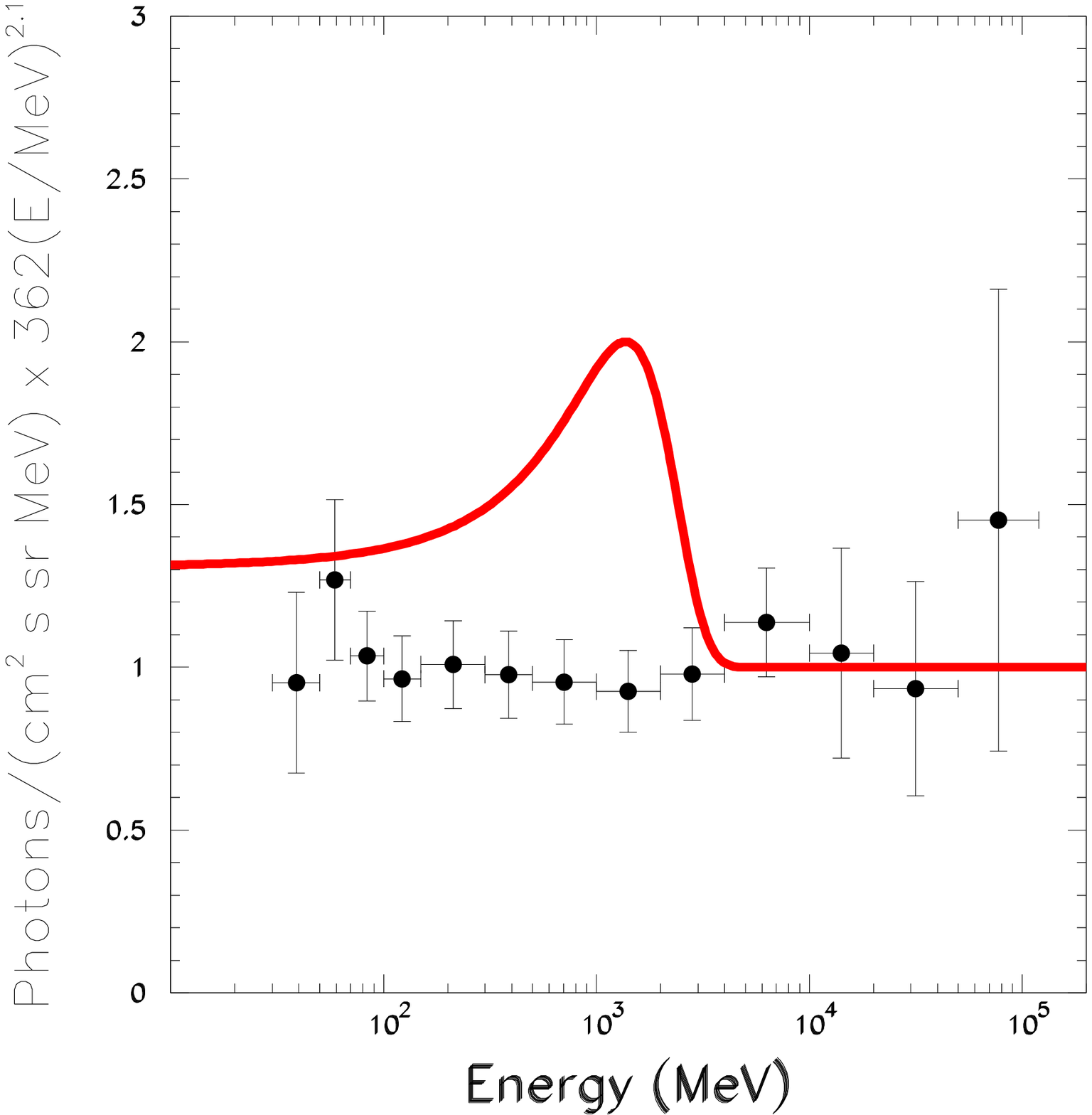,width=12.cm, height=9.cm}
\epsfig{file=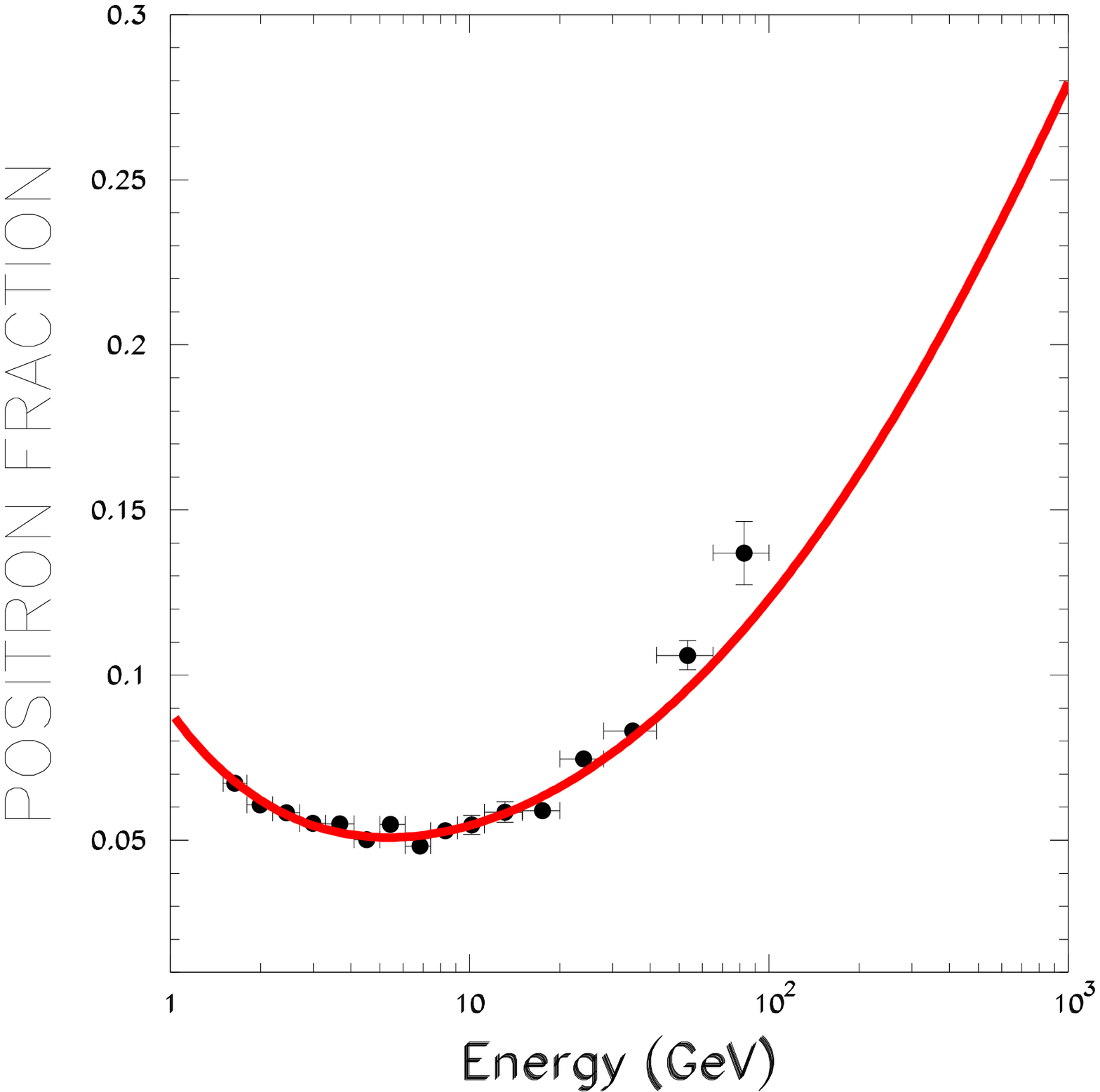,width=12.cm,height=9.cm}
}
\caption{ {\bf Top (a):} 
Comparison between the spectrum of the extragalactic GBR measured
by EGRET~\cite{Sreek} and a GBR spectrum which is produced by ICS of MBR
photons in external galaxies by a universal power-law spectrum of high
energy CR electrons, $dn_e/dE\!\propto\! E^{-3.2}$,
plus a universal excess such as that measured by ATIC~\cite{ATIC} between  
300-800 GeV~\cite{DMDD}. Both spectra were divided by the best fitted 
power-law to the EGRET GBR spectrum.
{\bf Bottom (b)}: Comparison between the  positron fraction
measured with PAMELA~\cite{Cafana},\cite{Morselli},\cite{PAMELA} 
and that expected
from secondary production of electrons and positrons in the CR
sources and in the ISM~\cite{DMDD}.}
\label{DMGBR}
\end{figure}

Moreover, it was pointed out~\cite{DMDD}
that if the ATIC electron excess  was due to dark matter annihilation,
such an excess of Galactic cosmic ray electrons would have produced 
a detectable GeV excess in the diffuse Galactic GBR at large latitudes, 
while dark matter annihilation in external galaxies would have produced 
a detectable GeV excess in the diffuse extragalactic GBR at 
all latitudes, which was not observed by EGRET (Figs.~\ref{DMGBR}a,b).

{\bf After the Rencontre de Moriond it was shown that the ATIC excess is 
probably 
instrumental due to misidentified proton induced electron-like events in 
the ATIC detector by cosmic ray protons~\cite{Fa}. Moreover, the HESS 
collaboration reported a measurement of the cosmic-ray electron spectrum 
above 340 GeV which does not show the ATIC peak~\cite{eHESS}
and the LAT collaboration
reported a high precision measurement of the steeply falling cosmic ray 
electron spectrum between 20 GeV and 1 TeV which also 
does not show the prominent ATIC peak~\cite{eLAT}. 
The spectral index of the CR electrons (plus positrons) 
with energy below 1 TeV which was measured by 
HESS and by PAMELA is consistent with -3.2.  This index is suggested by 
the spectral index
2.1$\!\pm\!0.03$ of both the Galactic 
GBR at large latitudes and the extragalactic GBR (at all latitudes), which
were measured by EGRET. However a significantly different spectral index 
$-3.04$ was measured by LAT. Additional measurements by LAT and by
other experiments in space such as
PAMELA, and in particular experiments with  magnetic spectrometers 
such as AMS  in space and on high altitude
balloon experiments above the south pole, are highly desirable.} 

\subsubsection{\bf The PAMELA positron fraction:} 

In the standard leaky box models, CR sources accelerate primary cosmic ray 
nuclei and electrons while secondary electrons and positrons are produced 
by the decay of charged $\pi$'s and $K$'s produced in hadronic collisions 
of primary cosmic ray nuclei in the interstellar medium (ISM). The primary 
particles are injected with roughly the same energy spectrum 
$dn/dE\!\sim\! E^{\!-\!p_{inj}}$ with $p_{inj}\!\approx\! 2.2$, but the 
escape by diffusion from the Galaxy increases the spectral index of the 
primary CR nuclei to $p_{_N}\!\approx\!2.7$ while cooling by synchrotron 
radiation and inverse Compton scattering of background photons increases 
the spectral index of the primary CR electrons by one unit to 
$p_e\!\approx\! 3.2$. Because of Feynman scaling the secondary electrons 
and positrons, which are produced by CR interactions in 
the ISM, have a 
spectral index $p_{inj}\!\approx\! 2.7$, which increases to 
$p_e\!\sim\!3.7$ by cooling. Consequently, in the standard CR model the 
positron fraction decreases like $\!\sim\! E^{\!-\!0.5}$ at high energies 
(where solar modulation and geomagnetic effects are negligible). 
Contrary to this expectation the PAMELA satellite experiment has recently 
reported~\cite{Cafana},\cite{Morselli},\cite{PAMELA},  a dramatic rise 
in the positron fraction starting 
at 10 GeV and extending up to 100 GeV in complete disagreement with the 
standard 
cosmic ray model calculations~\cite{MS}. These observations have 
created much 
excitement 
and motivated many papers claiming that the observed rise is produced by 
the annihilation of dark matter particles. Other publications related the 
excess to a local enhancement of the flux of electrons and positrons due 
to nearby galactic sources of positrons and electrons such as 
pulsars~\cite{Ka} or to secondary production in the ISM by CRs from nearby 
sources such as supernova remnants in the nearest spiral arm~\cite{Piran}.

However, the rise of the positron fraction with increasing energy 
beyond 10 GeV may be 
entirely due to hadronic 
production of positrons (and electrons) in the cosmic ray 
sources~\cite{DMDD}: In fact, if Fermi acceleration of highly relativistic 
particles 
results in a universal 
power-law 
distribution of Lorentz factors of the accelerated particles, 
$dn/d\gamma\!\propto\! \gamma^{-p_{inj}}$, with an injection spectral 
index $p_{inj}\!\approx\! 2.2$, than the injected flux of high energy 
electrons 
is suppressed by a factor $(m_e/m_p)^{p_{inj}-1}\!\approx\!10^{-4}$ 
compared to that of protons at the same energy~\cite{DDCR},  which is much 
smaller 
than their observed ratio in the Galaxy. Cosmic ray nuclei, however, may 
encounter in/near source a total column density comparable to a mean free 
path for hadronic interactions during their acceleration and before being 
injected into the ISM. In that case, due to Feynman scaling, they generate 
an electron+positron spectrum identical to that of the CR protons but with 
a 
normalization which is larger by roughly two orders of magnitude than that 
of the primary Fermi accelerated electrons. The combination of Fermi 
acceleration of electrons and hadronic production of electrons and positrons 
in/near the CR sources 
plus hadronic production of electrons and 
positrons in the ISM 
can naturally explain the rise of the positron 
fraction beyond 10 GeV~\cite{DMDD}.

\begin{figure}[]
\centering
\vbox{
\epsfig{file=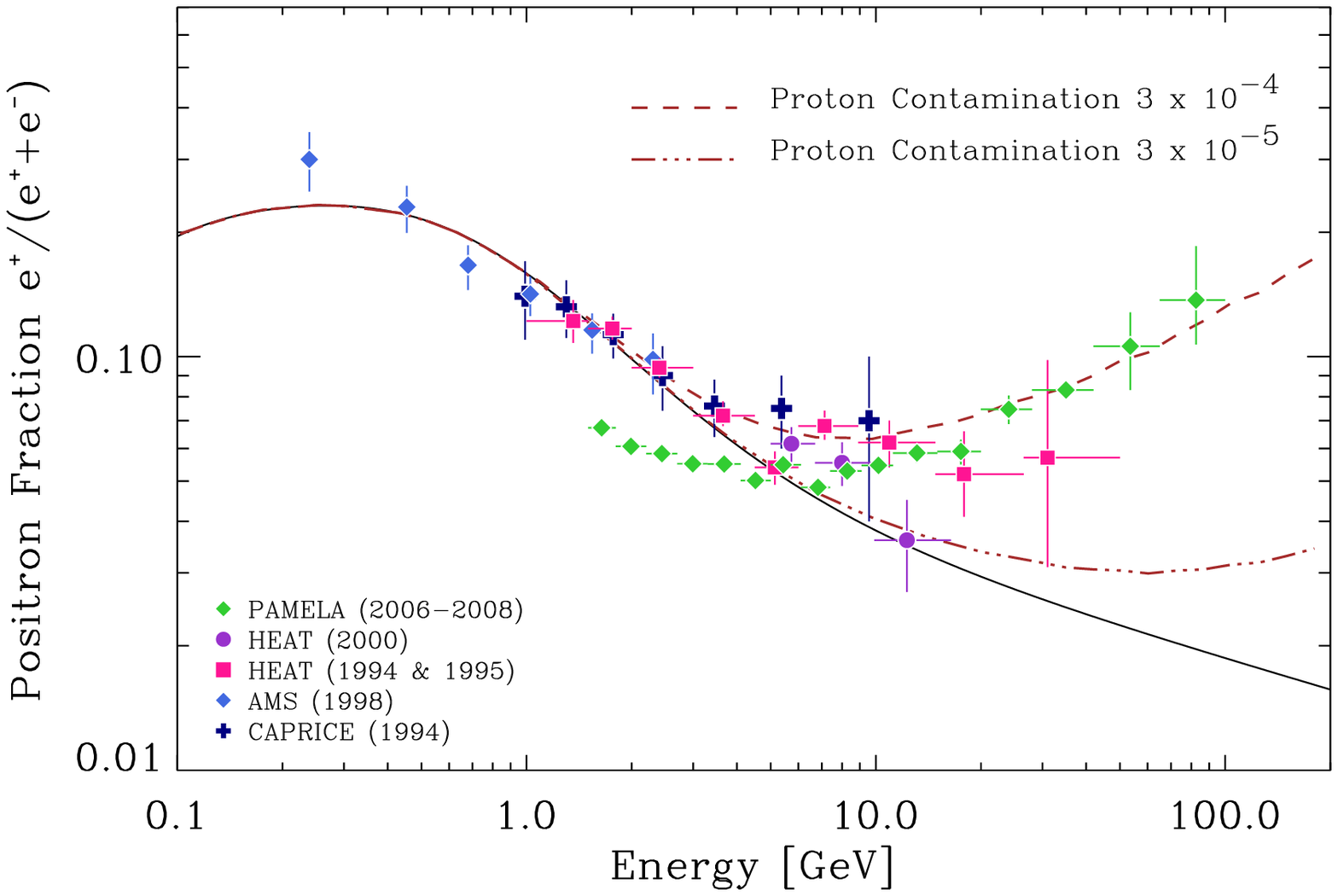 ,width=12.cm, height=9.cm}
\epsfig{file=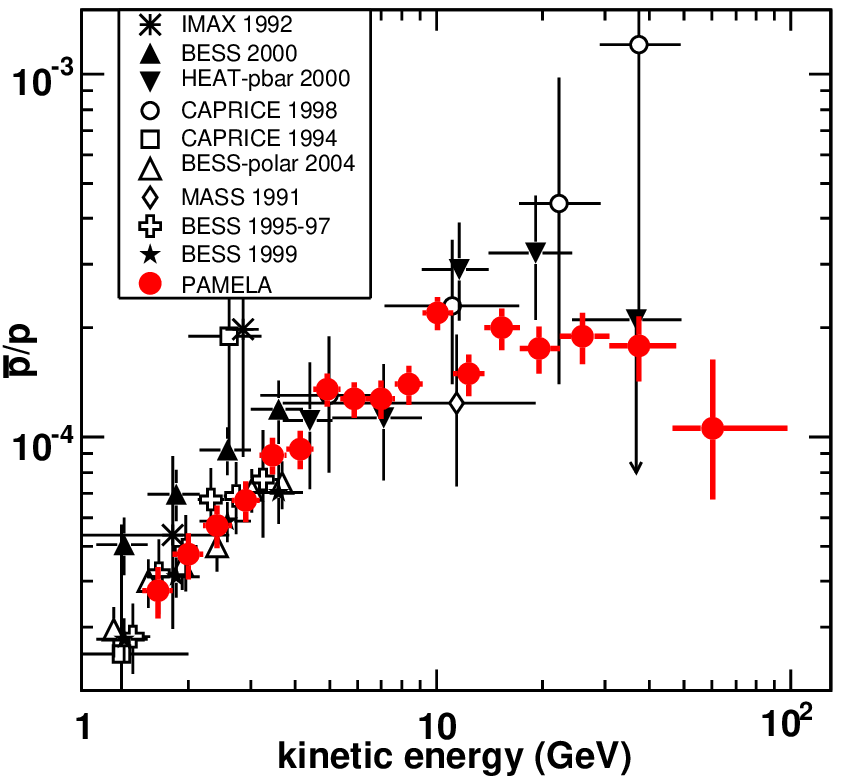,width=12.cm,height=9.cm}
}
\caption{ 
{\bf Top (a):} Recent measurements of the positron fraction overlaid with a the standard
leaky box model prediction~\cite{MS} of secondary production of cosmic-ray
positrons
in the ISM and the same prediction including residual proton
contamination~\cite{Schub}. Below 5 GeV solar modulation affects the
particle
intensities observed near Earth and may explain the discrepancy between
the PAMELA data and older measurements, obtained at distinctively
different solar epochs. In the region between 5 and 50 GeV measurements by
PAMELA are consistent with previous data from the HEAT experiment.
{\bf Bottom (b):} Comparison between the antiproton to proton 
ratio in Galactic cosmic rays as function of energy
as measured by PAMELA and by previous experiments.
The results of PAMELA cannot distinguish yet 
between a ratio decreasing 
with energy as expected from
secondary production of antiprotons in the ISM,  and roughly a 
constant ratio expected from secondary production in the CR sources.
} 
\label{pbarposPAMELA}
\end{figure}

Finally, despite of the above, caution must be applied also to the PAMELA 
results as emphasized in this Rencontre by M. Schubnell~\cite{Schub}: The 
intensity of cosmic-ray 
protons at 10 
GeV exceeds that of positrons by a factor of about $5\times 10^4$. 
Therefore a proton rejection of about $10^6$ is required if one wants to 
obtain a positron sample with less than 5\%. Furthermore, because the 
proton spectrum is much harder than the electron and positron spectra, the 
proton rejection has to improve with energy. In addition, any small amount 
of spillover from tails in lower energy bins can become 
problematic~\cite{Schub}. Fig.~\ref{pbarposPAMELA}a demonstrates that a 
proton 
contamination of 
$3\times 10^{-4}$ can explain the PAMELA positron fraction.

\subsubsection{\bf The PAMELA antiproton to proton ratio:} 

The recent measurements of the antiproton to proton ratio measured by 
PAMELA~\cite{Cafana},\cite{Morselli} agrees with that expected from 
secondary production in the ISM, but the measurements do not extend to 
high enough energy (see Fig.~\ref{pbarposPAMELA}b) where the energy 
dependence 
can distinguish between secondary production 
in the CR sources which yields a constant 
ratio and secondary production in the ISM that yields a ratio 
which decreases like  $E^{-0.5}$.

\section{High Energy Gamma Ray Astronomy}

\begin{figure}[]
\centering
\epsfig{file=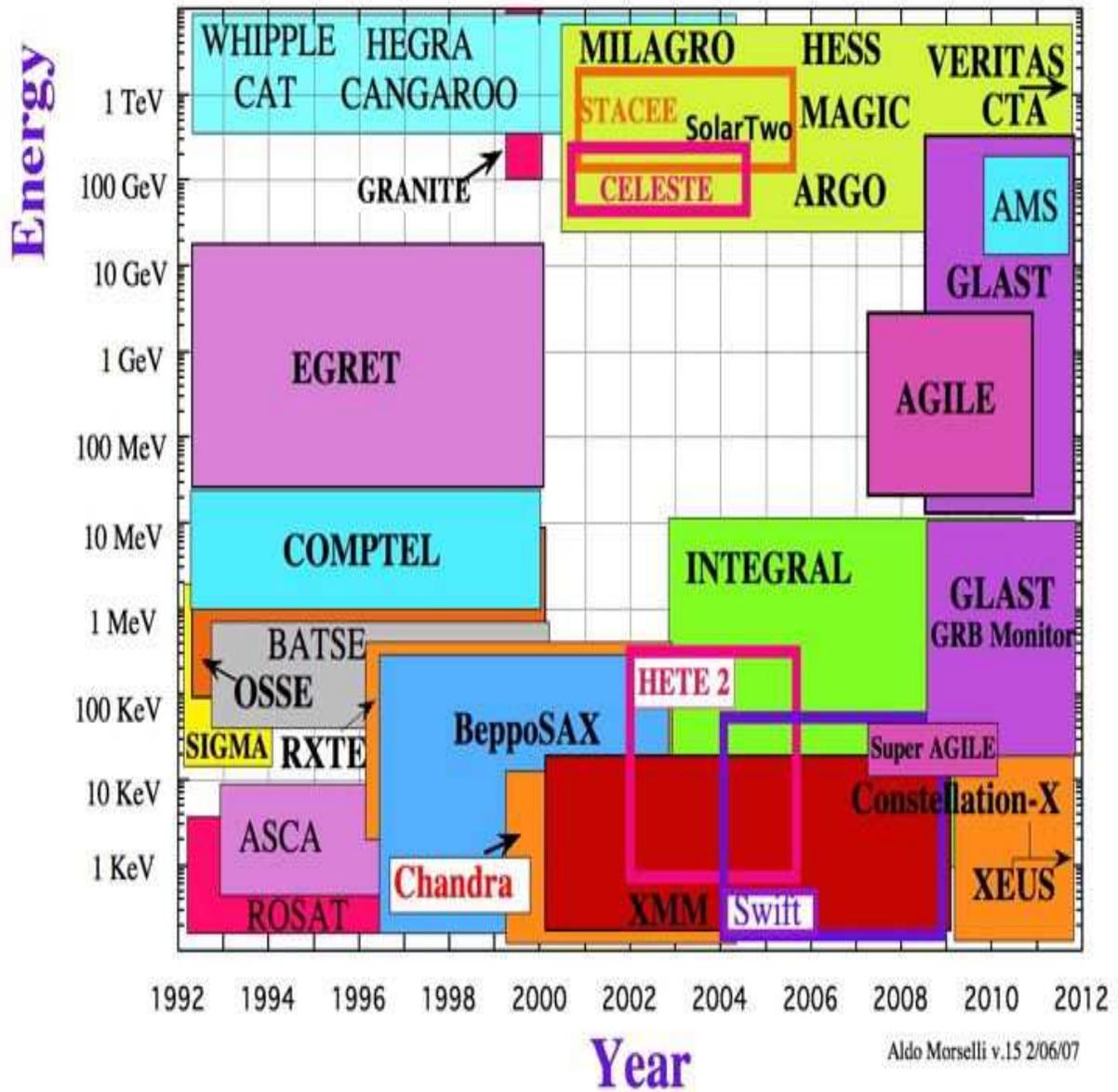,width=16.cm,height=16.cm}
\caption{The increasing energy range and sky coverage in the past 20 years
by water and air-shower Cherenkov telescopes and by gamma ray 
telescopes aboard satellites.}
\label{grtelescopes}
\end{figure}

The tremendous progress made in high energy gamma ray astronomy during the 
past two decades is due to many instruments with increasing sensitivity 
covering now the entire MeV-PeV energy range, as summarized in 
Fig.~\ref{grtelescopes} borrowed from Aldo Morselli.

This progress has culminated with the 
successful completion and operation of the large imaging Cherenkov 
telescope systems, HESS, MAGIC and VERITAS and the launch of the Fermi 
Gamma-ray Space Observatory on June 11, 2008 with its two main 
instruments, 
the Large Area Telescope (LAT) for all-sky survey studies of 
astrophysical and cosmological point and diffuse sources of high energy 
($30\!<E<\!300$ GeV) and Gamma-ray Burst Monitor (GBM) to study gamma-ray 
bursts. These studies led to an explosion of newly discovered 
Galactic and extragalactic sources.

Most of the 125 bright non-pulsar gamma ray sources detected by LAT at 
high latitude ($b\!>\!10^O$ in the first 3 months of operation are AGNs 
(57 FSRQ, 42 BLLac, 6 of uncertain class and 2 radio 
galaxies)~\cite{Lott}. The Galactic gamma ray sources include 13 new 
pulsars~\cite{Giordano} (radio-quiet pulsars, young radio pulsars and 
millisecond pulsars), pulsar wind nebulae~\cite{Grondin}(PWNe), supernova 
remnants, molecular clouds, X-ray binaries~\cite{Hill}, Wolf-Rayet stars, 
OB associations, open clusters and globular clusters~\cite{Smith}.

\subsection{\bf High energy gamma ray astronomy and the origin of 
Galactic CRs}

In 1934, Baade and Zwicky proposed that supernovae are the main sources of 
galactic CRs which were first discovered by Hess in 1912. Today diffusive 
shock acceleration in the blast wave driven into the ISM by a supernova 
shell is the most popular model for the origin of galactic cosmic rays. 
Despite the general consensus and exciting recent results, the origin of 
these particles is still debated and an unambiguous and conclusive proof 
of the supernova remnant hypothesis is still missing. In particular, the 
recent detection of a number of supernova remnants in TeV gamma rays by 
HESS, MAGIC and VERITAS still does not constitute a conclusive proof that 
galactic cosmic rays nuclei with energies below the cosmic ray knee are 
accelerated mainly in supernova remnants (SNRs). In particular, it was 
found that it is difficult to disentangle the hadronic and leptonic 
contributions to the observed gamma ray emission (for an excellent review 
see~\cite{Ga}).
 
In some shell SNRs such as RX J1713.7-3946 and Vela Junior the non-thermal 
synchrotron emission exhibits a striking morphological similarity with the 
TeV gamma ray image. Such a correlation is naturally expected in leptonic 
models, where both X-rays and gamma rays are emitted by the same 
population of electrons via synchrotron and inverse Compton scattering, 
respectively. Although the correlation can be accommodated also within 
hadronic models if most of the gamma ray emission is through $\pi^0$ decay 
and the X-ray emission is the result of synchrotron emission from 
secondary electrons from $\pi^{\pm}$ decay. In such a scenario the energy 
flux in TeV gamma rays must exceed that in X-rays since the electrons from 
$\pi^{\pm}\!\rightarrow\! \mu^{\pm}\!\rightarrow\! e^{\pm}$ decay carry 
less energy than the $\gamma$'s from $\pi^0$ decay, while the opposite is 
observed in RX J1713.7-3946.  But, the assumed synchrotron radiation from 
secondary electrons plus positrons may not be the correct origin of the 
X-ray emission from RX J1713-394 (e.g. bremsstrahlung from ISM protons 
which enter the SN shell rest-frame with $\!\sim\!$ 200 keV kinetic 
energy). In fact, the gamma ray spectrum that was measured from this SNR 
by HESS up to almost 100 TeV has a knee (or an exponential cutoff) around 
$E\!\sim\!5$ Tev which suggest that protons are accelerated in RX 
J1713.7-3946 up to the CR knee energy around 2 PeV:  At 2 PeV the mean 
charge multiplicity (mostly pions) in pp collisions is around 50 and that 
of the $\pi^0$'s is about 25. Pions carry about 35\% of the incident 
proton energy and about 1/3 of that energy is carried by $\pi^0$'s. 
Consequently, the typical energy of photons from the decay of $\pi^0$ 
produced by 2 PeV protons in pp collisions is roughly 5 TeV.

However, the safest way of proving or rejecting acceleration of CR nuclei 
in RX J1713.7-3946 (and in SNRs in general) is to search for neutrinos 
produced in the decay of charged pions (by stacking all the neutrino 
events from the direction of known SNRs).

\subsection{\bf High energy gamma ray emission in GRBs}

During nearly 20 years of observations the Burst And Transient Source 
Experiment (BATSE) on board the Compton Gamma Ray Observatory (CGRO), has 
detected and measured light curves and spectra in the keV-MeV range of 
several thousands Gamma Ray Bursts (GRBs). Higher-energy observations with 
its EGRET instrument aboard CGRO were limited to those GRBs which happened 
to be in its narrower field of view. Its large calorimeter measured the 
light-curves and spectra of several GRBs in the 1-200 MeV energy range. 
Seven GRBs were detected also with the EGRET spark chamber, sensitive in 
the energy range 30 MeV - 10 GeV. The EGRET detections indicated that the 
spectrum of bright GRBs extends at least out to 1 GeV, with no evidence 
for a spectral cut-off (see, e.g., Dingus 2001, and references therein). 
However, a few GRBs, such as 940217~\cite{Hurley} and 
941017~\cite{Gonzalez} showed evidence for a high energy component in the 
GRB pulses which begins significantly after the beginning of the keV- MeV 
pulse and has a slower temporal decay than that of the keV-MeV emission, 
suggesting that the high-energy emission, at least in some cases, is not a 
simple extension of the main component, but originates from a different 
emission mechanism and/or region. This has been confirmed recently by 
observations of high energy photons from several GRBs with the Fermi 
LAT~\cite{Baldini},\cite{Pelassa},\cite{Granot}, and 
AGILE~\cite{AGILE}.  
However, the flux levels of TeV gamma rays from a couple of GRBs which 
were inferred from ground level measurements of atmospheric showers were 
not confirmed by HESS with its high sensitivity array which produced 
upper 
limits much smaller than the flux levels predicted by standard fireball 
models where TeV photons are produced by inverse Compton scattering, decay 
of $\pi^0$'s from proton-gamma collisions and synchrotron radiation 
from UHE protons.
  
\begin{figure}[]
\centering
\vbox{
\hbox{
\epsfig{file=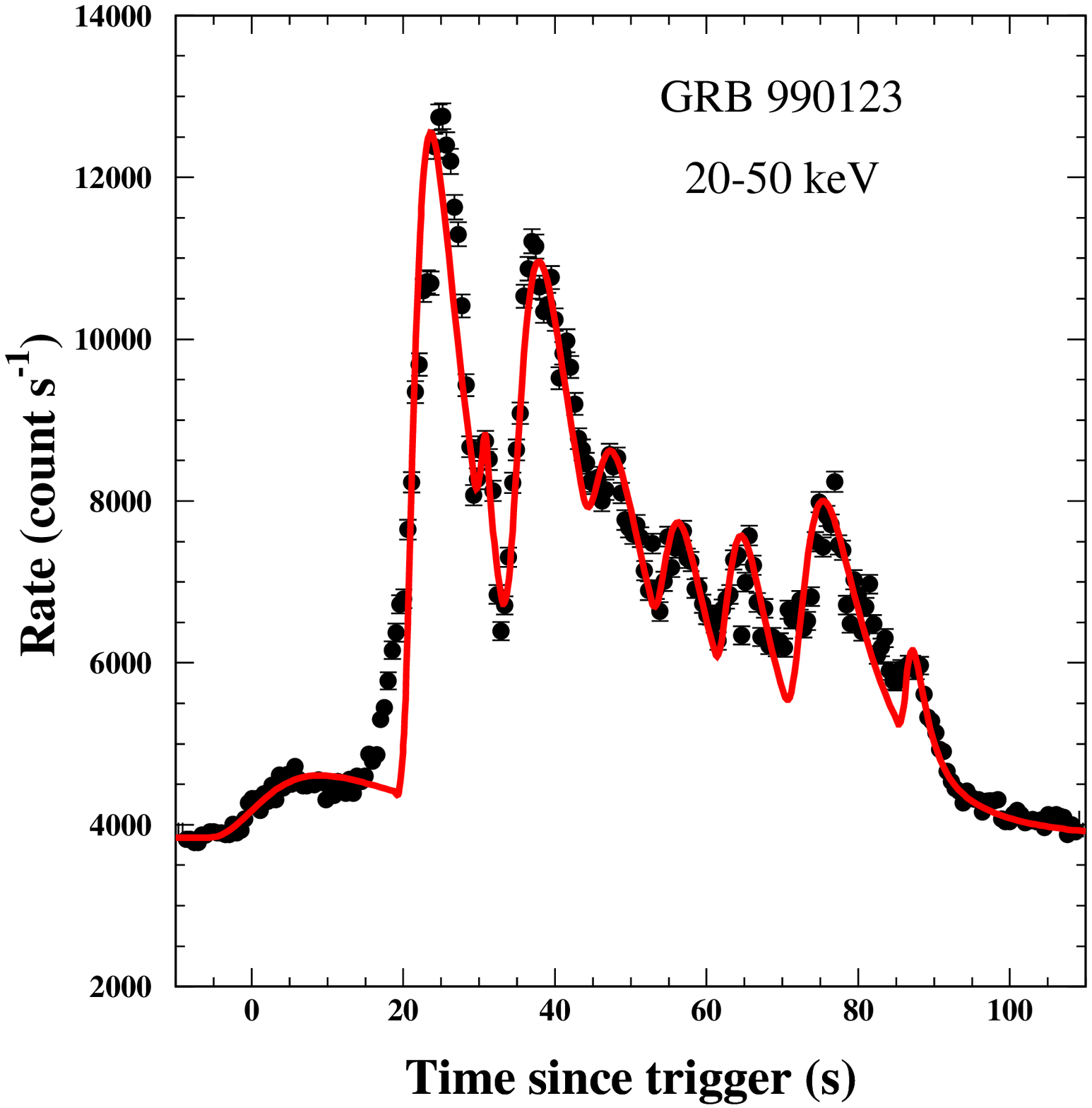,width=8.cm,height=8.cm}
\epsfig{file=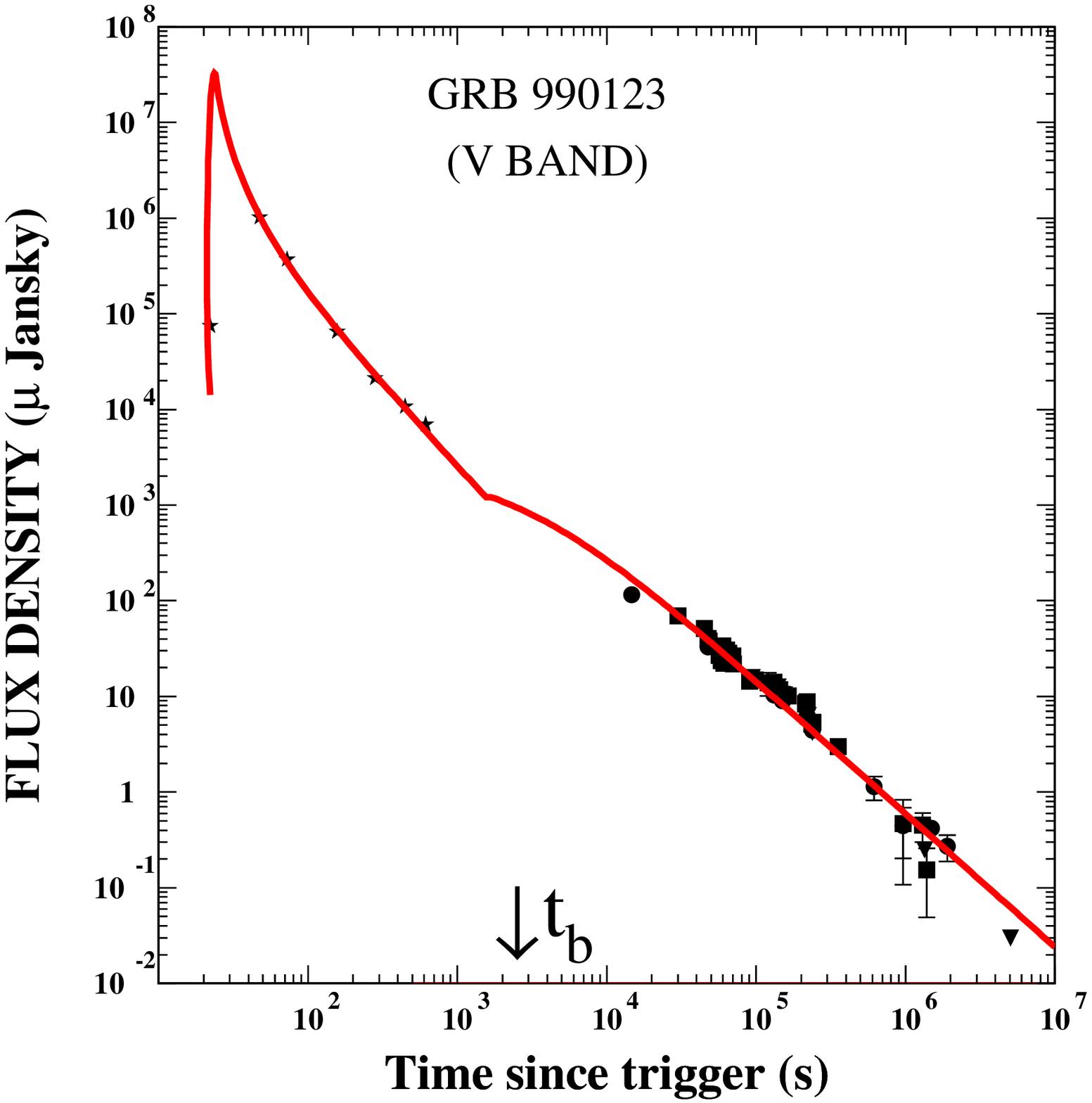,width=8.cm,height=8.cm}
}}
\caption{ {\bf Top (a):} Comparison between the 20-50 keV BATSE lightcurve
of GRB990123~\cite{Briggs} and its CB model description~\cite{DD2GRB}.
The sub-pulse superimposed on the
decaying tail of the three major pulses may be due to
the crossing of the 3 leading CBs through two successive wind layers
(2 separate pre-supernova mass ejections by the progenitor star)
rather than by 3 additional CBs.
{\bf Bottom (b):} The entire $V$ band lightcurve of GRB 990123
and its CB model description as a synchrotron emission 
from the collision of the jet of CBs with a wind (with a density
profile $n\!\propto\!1/(r\!-\!r_0)^2$ for $r\!>\!r_0$)  
overtaken  by a constant ISM
density around an observer time $t\!=\!1000$ s.
The `prompt' (early-time)  $V$ band lightcurve was measured with 
ROTSE~\cite{Akerlof} which did not resolve it into individual 
peaks. It shows a 
time lag of several seconds of the prompt optical emission relative to 
the  prompt keV-MeV emission.}
\label{GRB990123}
\end{figure}

Not only the observed flux levels but also the spectral and temporal 
behaviour of the high energy emission are not those predicted by the 
popular fireball (FB) models of GRBs. This is not completely surprising in 
view of the fact that the rich and accurate data, which have been 
accumulated in recent years from space-based observations with Swift and 
ground based observations with robotic telescopes, have already challenged 
the prevailing popular views on GRBs: Synchrotron radiation (SR) 
cannot explain simultaneously their prompt optical emission and their hard 
X-ray and gamma-ray emission which were well measured in some bright GRBs 
such as 990123 and 080319B (Figs.~\ref{GRB990123},\ref{GRB080319B}).
The prompt hard X-ray and gamma-ray pulses 
cannot be explained by synchrotron radiation from internal shocks 
generated by collisions between conical shells. Neither can SR explain 
their typical energy, spectrum, spectral evolution, pulse-shape, rapid 
spectral softening during their fast decay phase and the established 
correlations between various observables. Moreover, contrary to the 
predictions of the FB model, the broadband afterglows of GRBs are highly 
chromatic at early times, the brightest GRBs do not show jet breaks, and 
in canonical GRBs where breaks are present, they are usually chromatic and 
do not satisfy the closure relations expected from FB model jet breaks. In 
spite of all the above, the GRB community is not so critical and many 
authors believe that the GRB data require only some modifications of the 
standard FB model in order to accommodate the observations. Other authors 
simply ignore the failures of the FB model and continue the interpretation 
of observations with the FB model hypotheses (colliding conical shells, 
internal and external shocks, forward and reverse shocks, continuous 
energy injection, refreshed shocks) and parametrize the data with freely 
adopted formulae (segmented power laws, exponential-to power-law 
components) which were never derived explicitly from any underlying 
physical assumptions. 

In spite of the above, not all authors are so critical and they believe 
that the GRB data require only some modifications of the standard FB model 
in order to accommodate the observations. Many papers, including some 
presented at this Rencontre seem to ignore the failures of the FB model 
and continue to interpret the observations with the FB model hypotheses 
and parametrize the data with freely adopted formulae.

The situation of the cannonball (CB) model of GRBs is entirely different. 
In a series of publications, which were largely ignored by the rest of the 
GRB community, it was demonstrated repeatedly that the model correctly 
predicted the main observed properties of GRBs and reproduces successfully 
the diverse broad-band light-curves of both long GRBs~\cite{DDD1} and 
short hard bursts (SHBs)~\cite{DDD2}. In fact since the discovery of GRBs 
in 1967
and the beginning of the GRB debate, the majority view  
on key GRB issues initially was always wrong, while a minority view 
turned out to be the right one,  as demonstrated  in Table I
where the `correct view' is indicated by  bold letters.

In the CB model, a highly relativistic jet of plasmoids (CBs) from the 
central engine first encounters a cavity produced by the wind/ejecta blown 
by the progenitor star (SN-GRBs) or by a companion star or an accretion 
disk in abinary system (SHBs) and filled up with quasi isotropic radiation 
(glory) emitted/scattered by by the wind/mass ejecta prior to the GRB. The 
prompt gamma-ray and X-ray emission is dominated by inverse Compton 
scattering (ICS) of this glory light.  A simultaneous broad band 
synchrotron radiation (SR) and inverse Compton scattering of this 
radiation to much higher energies begin slightly after the CBs have swept 
in enough electrons and ionized nuclei of the ejecta/wind in front of 
them, isotropized them and Fermi accelerated them and the knocked-on 
(Bethe-Bloch) electrons and nuclei in the CBs to high energy by their 
turbulent magnetic fields. SR from these electrons dominates the optical 
radiation, while ICS of these SR photons (SSC) produces high energy 
photons with an energy flux density that extends beyond TeV. Production of 
$\pi^0$'s in collisions between the Fermi accelerated nuclei and the 
ambient matter in the CBs and the wind produces a power-law distribution 
of high energy photons which extends to much higher energies. The same 
mechanisms can produce also the observed high energy emission from short 
hard bursts (SHBs). Like for blazars, the observed flux of high energy 
photons from ordinary GRBs and SHBs is suppressed significantly at TeV 
energies by pair production in the IGM, while in the energy range covered 
by LAT, the absorption of photons by the extragalactic background light 
can be negligible.

\section{High energy gamma ray astronomy, UHECRs and the extragalactic 
background light}

\begin{figure}[]
\centering
\epsfig{file=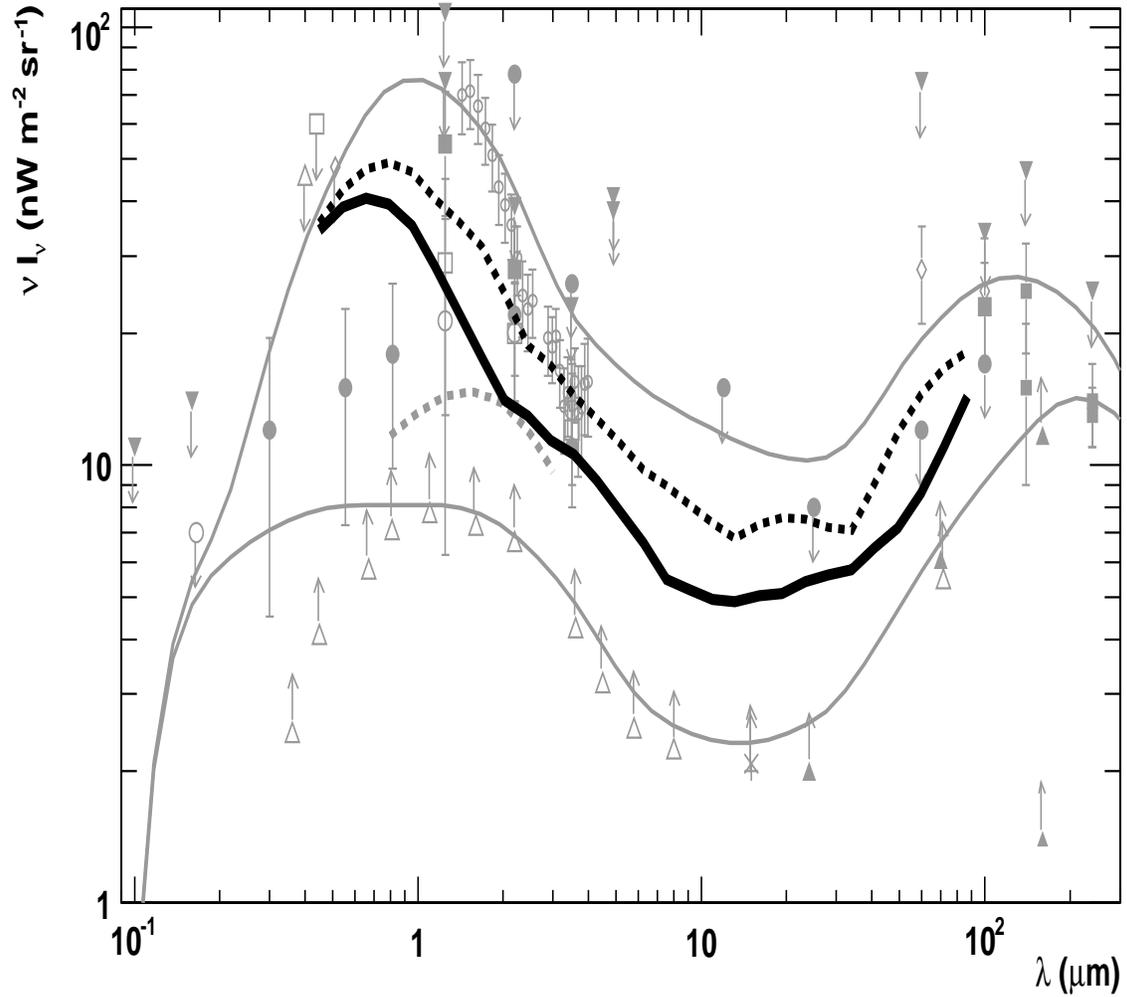,width=16.cm,height=14.cm}
\caption{Limits and estimates of the spectrum of the
extragalactic background light (EBL) as extracted from different
measurements and theoretical models prior to the detection of
the blazar 3C279 by MAGIC in TeV gamma rays~\cite{Mazin}.}
\label{EBL}
\end{figure}

Pair production in collisions of high energy photons with extragalactic 
background light (EBL) from the far infrared strongly modifies the flux 
and spectrum of high energy (0.1-100 TeV) photons from distant point and 
diffuse sources. Measurements of these fluxes from various bright sources 
such as AGNs and GRBs as function of redshift can be used to test and 
constrain theoretical models of star and dust formation, structure 
formation in the early universe, astrophysical models of HE 
cosmic sources and 
photon-photon interaction at very high energies. Photodisinegration of 
UHECR nuclei in their collisions with EBL photons strongly affects their 
composition\cite{Puget}. TeV gamma rays from blazars have been used 
extensively to test the measurements and theoretical estimates of the EBL 
(see ~\cite{Mazin} and Fig.~\ref{EBL}), the strongest constraints come 
from the most distant blazar
3C279 at $z$=0.536, which has been detected by MAGIC~\cite{Lotto} in 
TeV gamma rays. Detection of a 13 GeV photon from
GRB 080916C 
with the Fermi LAT at redshift $z$=4.35 has also been used already 
to test different EBL
models~\cite{Abdo}.

\subsection{\bf HE gamma rays from extragalactic sources}

Despite the detection of a dozen of extragalactic 
blazars in TeV by HESS~\cite{Gerard}, MAGIC~\cite{Lotto} 
and VERITAS~\cite{Benbow}
and ten times more in GeV photons by Fermi LAT~\cite{Ballet},\cite{Lott} 
and despite the multi wavelength
campaigns (e.g. \cite{Sanchez} where a few of these extragalactic sources 
were observed 
simultaneously in the radio, optical, X-ray, GeV and TeV
bands, beside constraining some theoretical models, 
not much better understanding of how massive black holes
launch their mighty jets has been achieved. 
This is because of the complexity of the
black hole engine, the complexity of its environment, the complex
time variability of the observed emission 
and the very many adjustable parameters 
and assumptions in the theoretical models. Roughly, most observations are 
consistent with a leptonic SSC model   where 
synchrotron radiation 
from a population of Fermi accelerated  
electrons with a typical peak flux energy $E_{SR}$ 
suffers inverse Compton scattering by the same population of electrons.
The relativistic kinematics and the energy dependence of the 
Klein-Nishina cross section of ICS
produces a second peak at $E_{SSC}\!\approx\!(m_e\, c^2)^2\delta^2/ 
3\, E_{SR}\,(1+z)^2$ where $\delta$ is the Doppler factor of 
the Blazar's jet.

\begin{table}[t]
\caption{Evolution of the GRB debate}
\vspace{0.4cm}
\hskip -0.5cm
\begin{center}
\begin{tabular}{|c|c|c|l|}
\hline
& & &\\
Issue & Majority View & Minority View & Observational Proof (Year)\\   
\hline
Origin  &  Man Made  & {\bf Nature Made} & Vela Satellites (1967-1973) \\
& & &\\
Location & Solar System   & {\bf More Distant}  & Vela Satellites 
(1967-1973) \\
      & Galactic Disk & {\bf Cosmological}  &    CGRO (1992)\\
      & Galactic Halo & {\bf Distant Galaxies} & BeppoSAX+HST+GBTs (1997)\\
 & & &\\
Event & $n^*$-$n^*$ Merger &  {\bf SN Explosion} & BeppoSAX+HST+GBTs 
(1998-2003)\\
& & & \\
Source  & Relativistic Fireball & {\bf Relativistic Jet} & CGRO, 
BeppoSAX (1992-1999)\\ 
        & Collimated Fireball/  &{\bf Relativistic Jet} &  Swift+GBTs (2004-2009)\\   
           & Conical Jet         &              &  \\  
Prompt Radiation: & & &\\
keV-MeV & Synchrotron   & {\bf Inverse Compton} &  BeppoSAX, Swift 
                (1999-2009)\\
"Prompt Optical" & Reverse Shock & {\bf Synchrotron} & Robotic 
Telescopes 
(1999-2009)\\
Afterglow:  & & &\\      
Chromaticity & Achromatic  &  {\bf Chromatic}       &  Swift+Robotics+GBTs
(2004-2009)\\
Plateau phase  & Reenergization & {\bf Slow Deceleration} & Swift+GBTs
2004-2009\\
Jet Break when:  & $1/\Gamma_{jet}\!\approx\!\theta_{jet}$ & ${\bf 
\Delta M\!\approx\!M_0(jet)}$ & Swift+GBTs (2004-2009)\\
"Missing Break"  & Very Late Break& {\bf Very Early Break} & Swift+GBTs
(2004-2009) \\      
& & &\\
\hline
To be determined ? & & & Observational Proof by ? \\
\hline
& & & \\
Jet Geometry  & Conical Shells &  Cannonballs &   Swift,Fermi,HST,GBTs 
\\
Jet Composition & $e^+e^-$ plasma  & Ordinary Matter & Swift,Fermi, 
HST,GBTs  \\  
Beamed $E_\gamma$ &  $\sim 10^{51}$ erg  &~$\sim 10^{48}$ erg 
&Swift,Fermi,HST,GBTs\\ 

& & &\\
Source & Hypernova &  Normal SNIb/c  & Integral, Swift, Fermi, HST,GBTs  \\
   & (Rare SNIb/c) &  Most SNIb/c& Integral, Swift,Fermi,HST,GBTs  \\
Radiations:&&&\\
keV-MeV $\gamma$'s  &   SSC of ? & ICS of Glory Light &  Swift,Fermi,GBTs \\ 
HE $\gamma$'s  &     ?   & SSC + $pp\!\rightarrow\!\pi^o$ & 
LAT,HESS,MAGIC,VERITAS,PAO\\
HE Neutrinos & Detectable by& Not Detectable by:& ICECUBE,ANTARES,PAO \\ 
& & & \\
Remnant  &  BH, Magnetar &  $n^*$, BH & Swift,Fermi,HST,GBTs \\
& & &\\
XRFs &  Not GRBs & Far off-axis GRBs & Swift,Fermi,HST,GBTs\\
& & &\\                 

\hline
\end{tabular}
\end{center}
\end{table}

\section*{Acknowledgments} The author would like to thank all the speakers 
at the 44th Rencontre De Moriond on High Energy Phenomena in the Universe, 
the scientific organizing committee for a most interesting program, and in 
particular Jean Tran Thanh Van and Kim who step aside after having 
initiated and organized the past 44 Rencontres de Moriond in order to 
promote scientific collaboration, scientific exchange and spread of 
scientific knowledge beyond borders and ideological and racial barriers.
John Belz, Marusa Bradac, Francesco Cafana, Gudlaugur Johannesson, 
Daniel Mazin, Aldo Morselli and Michael Schubnell are gratefully 
acknowledged for supplying original figures for this summary.

\newpage

\newpage

\newpage

\newpage

\newpage
 
\newpage

\newpage

\newpage

\newpage
\begin{figure}[]
\centering
\vbox{
\epsfig{file=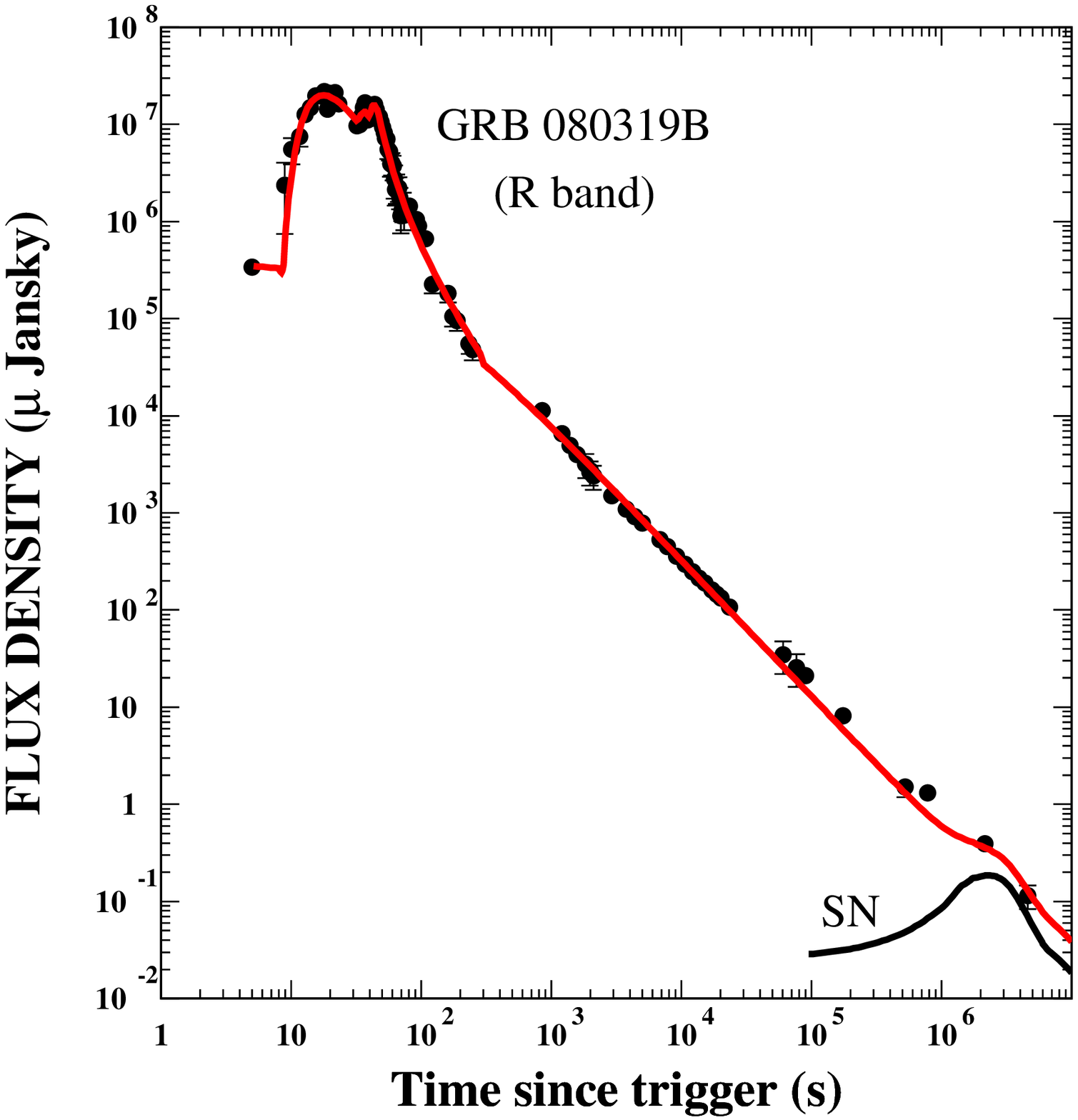,width=8.cm,height=8.cm}
\epsfig{file=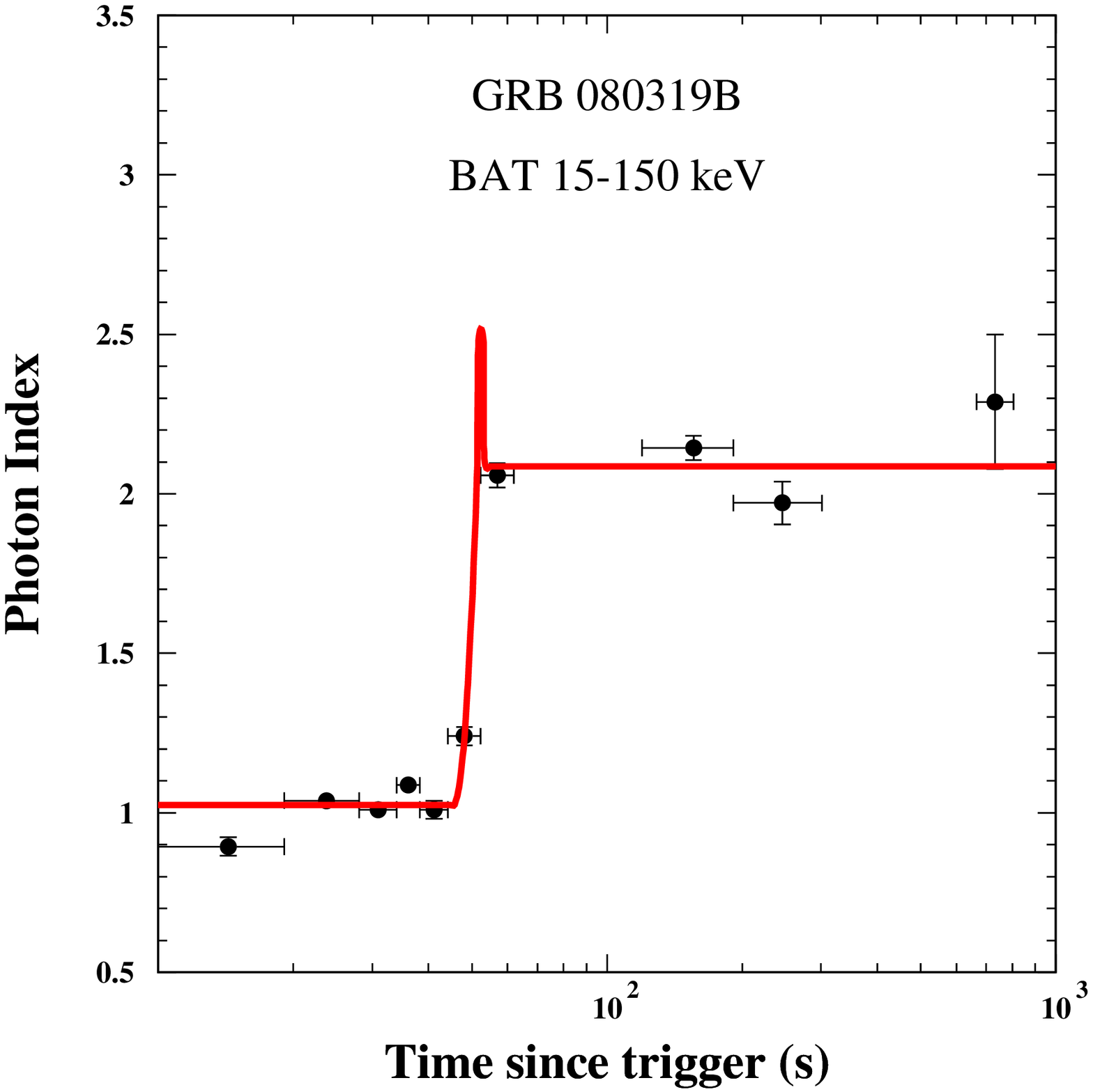,width=8.cm,height=8.cm}
}
\caption{{\bf Top (a):} The entire $R$-band (and $V$ band renormalized
to the $R$ band) lightcurve of GRB080319~\cite{Racusin}  and 
its CB
model description as synchrotron radiation from an initially expanding 3 
leading CBs which 
merged into a
single CB by the time they met the constant density ISM at the 
end 
of the
prompt ICS emission of gamma-rays and hard X-rays around 300 s (observer
time). Note that the prompt optical emission began about 10 seconds after 
the beginning of the keV-MeV emission.
Shown also is the 
contribution
to the $R$-band afterglow from SN akin to SN1998bw displaced to the GRB site.
{\bf Bottom (b)}: The mean photon spectral index in the 15-150
keV band as measured with the Swift broad alert telescope 
(BAT)~\cite{Racusin} 
and its CB model prediction. In
the CB model, the prompt emission is dominated by ICS of thin
bremsstrahlung with a typical $\Gamma\!\approx \!1$, which increases
rapidly during the fast decay phase of the prompt emission
and becomes $\!\approx\!2.1$
the typical value predicted by the CB model as soon as SR dominates the 
X-ray emission~\cite{DDD4}. 
}
\label{GRBO80319B}
\end{figure}

\newpage

\end{document}